\documentclass[times,twocolumn,final]{elsarticle}
\usepackage{medima}
\usepackage{framed,multirow}
\usepackage{makecell}
\usepackage{amssymb}
\usepackage{latexsym}

\usepackage{url}
\usepackage{xcolor}

\usepackage{hyperref}
\usepackage{array}
\usepackage[fleqn]{amsmath}

\definecolor{newcolor}{rgb}{.8,.349,.1}

\journal{Medical Image Analysis}

\begin{document}

\verso{Jihoon Cho \textit{et~al.}}

\begin{frontmatter}

\title{Principled Feature Disentanglement for High-Fidelity Unified Brain MRI Synthesis}%
%disentanglement, complementary, HF-GAN
% \tnotetext[tnote1]{This is an example for title footnote coding.}

\author[1,2]{Jihoon \snm{Cho}}
\author[3]{Jonghye \snm{Woo}}
\author[1]{Jinah \snm{Park}\corref{cor1}}
\cortext[cor1]{Corresponding author at: School of Computing, Korea Advanced Institute of Science and Technology, Daejeon, Republic of Korea. \\
  \textit{E-mail address}: jinahpark@kaist.ac.kr (J. Park).}

\address[1]{School of Computing, Korea Advanced Institute of Science and Technology, Daejeon 34141, Republic of Korea}
\address[2]{Samsung Research, Seoul 06765, Republic of Korea}
\address[3]{Gordon Center for Medical Imaging, Massachusetts
General Hospital and Harvard Medical School, Boston, MA 02114, USA}

\received{20 October 2025}
\finalform{-}
\accepted{-}
\availableonline{-}
\communicated{-}

\begin{abstract}
%%%
Multisequence Magnetic Resonance Imaging (MRI) provides a more reliable diagnosis in clinical applications through complementary information across sequences. However, in practice, the absence of certain MR sequences is a common problem that can lead to inconsistent analysis results. In this work, we propose a novel unified framework for synthesizing multisequence MR images, called hybrid-fusion GAN (HF-GAN). The fundamental mechanism of this work is principled feature disentanglement, which aligns the design of the architecture with the complexity of the features. A powerful many-to-one stream is constructed for the extraction of complex complementary features, while utilizing parallel, one-to-one streams to process modality-specific information. These disentangled features are dynamically integrated into a common latent space by a channel attention-based fusion module (CAFF) and then transformed via a modality infuser to generate the target sequence. We validated our framework on public datasets of both healthy and pathological brain MRI. Quantitative and qualitative results show that HF-GAN achieves state-of-the-art performance, with our 2D slice-based framework notably outperforming a leading 3D volumetric model. Furthermore, the utilization of HF-GAN for data imputation substantially improves the performance of the downstream brain tumor segmentation task, demonstrating its clinical relevance. 
% The concept of HF-GAN for feature disentanglement offers a robust and effective solution for high-fidelity MRI synthesis, demonstrating clear superiority in both quantitative evaluations and clinical applications.

%%%%
\end{abstract}

\begin{keyword}
\MSC 41A05\sep 41A10\sep 65D05\sep 65D17
%% Keywords
\KWD Medical Image Synthesis \sep Feature Disentanglement \sep Data Imputation \sep Multisequence MRI
\end{keyword}

\end{frontmatter}

%\linenumbers

\section{Introduction}

Magnetic Resonance Imaging (MRI) is highly effective in extracting crucial information from soft tissue regions, making it a widely used imaging modality in clinical diagnosis and treatment. By leveraging the intrinsic MR properties of tissues---such as proton density, and T1 / T2 relaxation times---and the properties of externally applied excitation, we can acquire various MRI sequences, where each sequence exhibits distinct characteristics, particularly yielding unique contrasts between soft tissues. Considering that the significance of the features varies with the MR sequence, the use of a combination of sequences yields superior results compared to processing a single sequence~\citep{menze2014multimodal, schouten2016combining}. In practice, however, acquiring a complete set of multisequence MRI scans can be challenging due to various practical issues, including extended scan duration, patient-induced motion artifacts, and protocol variability across sites and scanners. Incomplete MR sequences, which lack essential information, lead to inconsistent analysis results, particularly for computer-aided diagnosis methods using deep learning that are sensitive to data distribution~\citep{chan2020computer}. However, re-scanning to recover the missing sequences is typically infeasible because of cost, scheduling constraints, and added patient burden.

\begin{figure*}[!t]
\begin{center}
\includegraphics[width=1.0\textwidth]{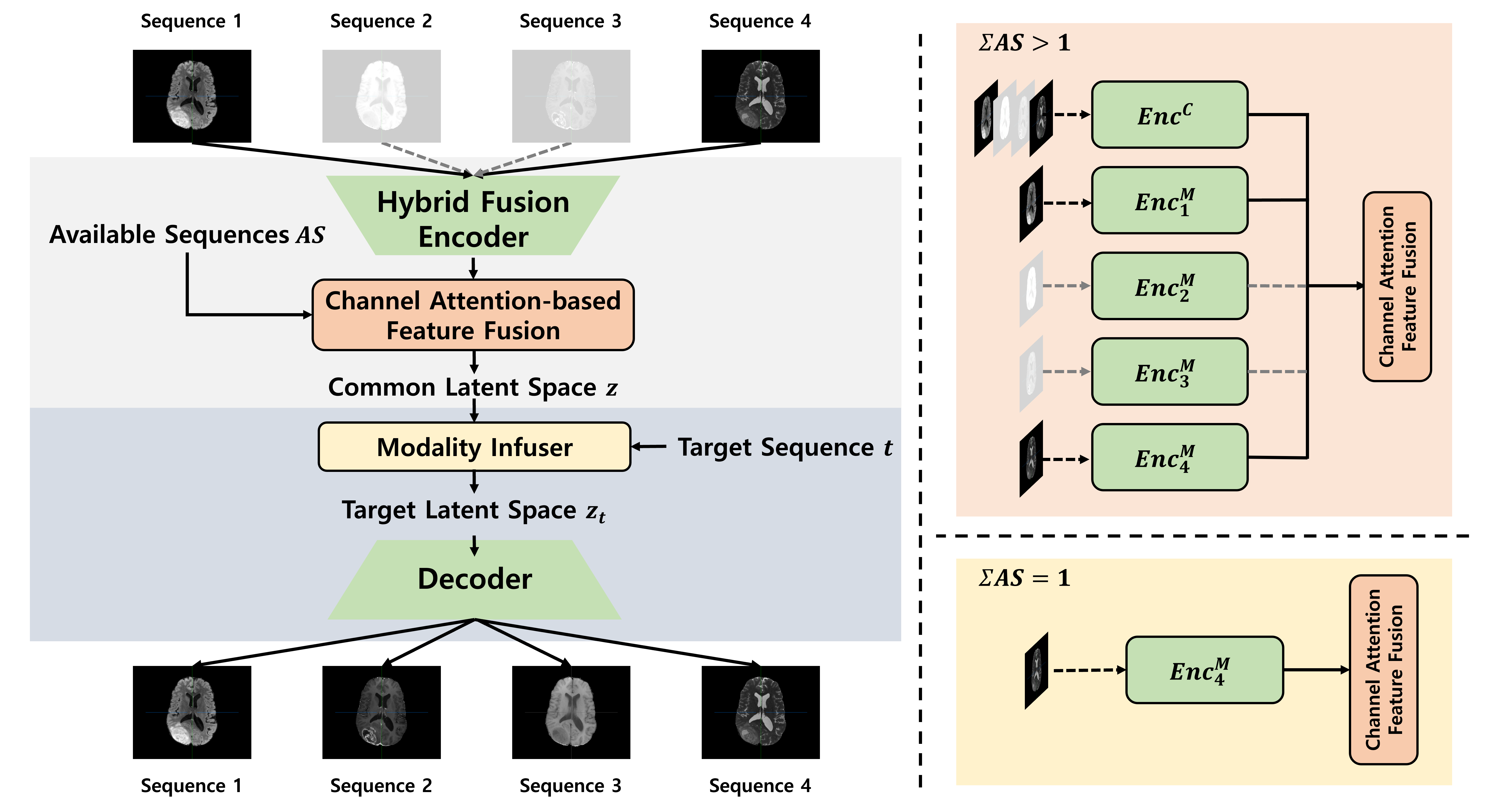}
\end{center} 
\caption{Illustration of our framework for synthesizing missing MR sequences. (left) Only accessible MR sequences are used to project into a common latent space. The projected feature representations are converted into the target latent space using the modality infuser, and a decoder generates target MR sequences based on the latent space of the feature representation. (right) If there are multiple accessible MR sequences, the complementary features extracted from the early fusion encoder are used.} 
\label{fig:overview}
\end{figure*} 

A common approach to address data imputation in medical imaging is to synthesize missing MR sequences based on available MR sequences. As deep learning-based approaches have advanced in medical imaging, MRI synthesis methods have also been translated into learning-based synthesis methods, such as convolutional neural networks (CNNs) and transformers~\citep{chartsias2017multimodal, dar2019image, sevetlidis2016whole, yurt2021mustgan, zhou2020hi} from traditional approaches~\citep{lee2017multi, jog2017random, ye2013modality}. To take advantage of CNNs, initial research focused on training the relationships between MR sequences by embedding them into a shared latent space~\citep{chartsias2017multimodal}, exploring multi-level interactions among sequences~\citep{zhou2020hi}, and constructing additional architectures to extract complementary information~\citep{yurt2021mustgan}. However, most of these methods are designed for fixed source-target sequence pairs, making them impractical for data imputation tasks in real-world scenarios.

Recent research has increasingly focused on unified approaches that can handle any combination of missing sequences within a single model~\citep{cho2024disentangled, dalmaz2022resvit, sharma2019missing, liu2021unified, liu2023one, yang2021unified}. However, information loss remains a challenge. For example, some frameworks may yield suboptimal performance, either because their single-input design prevents the full use of available MR data~\citep{cho2024disentangled, liu2021unified, yang2021unified}, or because their multi-input structure fails to create a clean latent space, allowing it to be corrupted by non-anatomical background features. Fully convolutional networks (FCN) also have fundamental limitations in learning long-range dependencies, which can be obstacles to their understanding of the overall structure~\citep{sharma2019missing}. Even with the application of transformer architecture that can capture long-range dependencies, the use of a single multichannel encoder~\citep{dalmaz2022resvit} or modality-specific encoders~\citep{liu2023one} alone is inadequate for capturing cross-sequence complementary information. Furthermore, these methods do not guarantee that their latent space is sufficiently robust to handle the complexity of all $2^N-2$ missing-sequence scenarios for $N$ MR sequences.

To address these challenges, in this work, we propose a novel unified framework for synthesizing missing MR sequences, called hybrid-fusion GAN (HF-GAN). HF-GAN can handle all missing-sequence combinations with a single network, fully exploiting the information available from MR sequences. To synthesize the missing sequence, HF-GAN employs a hybrid-fusion encoder~\citep{cho2023HFTrans} to separately extract complementary and modality-specific information. We achieve feature disentanglement by aligning architectural design with feature complexity. The extracted features are then projected into a common latent space to facilitate the sharing of essential MR sequence components through a channel attention-based feature fusion module. Subsequently, the feature representations in the common latent space are transformed into the target latent space using a modality infuser~\citep{cho2024disentangled} that employs consecutive self-attention mechanisms, and finally, the target MR sequence is synthesized using the CNN decoder. We evaluated our framework on multisequence MRI datasets of healthy brains and brains with tumors. The experimental results show state-of-the-art quantitative and qualitative performance. Furthermore, extensive analysis demonstrates that our approach can disentangle the latent space and extract complementary information within sequences. We also provide an ablation study validating each module’s contribution and a straightforward 3D extension using a simple yet effective module.

An earlier version of this work has appeared in ~\cite{cho2024disentangled}. Built upon that work, this paper introduces the following improvements. 
\begin{itemize}
    \item We extend the unified one-to-one translation method to a many-to-many synthesis method, utilizing all accessible MR images while maintaining a unified approach.
    \item We introduce a hybrid-fusion encoder designed to ensure the extraction of complementary information, along with a channel attention-based feature fusion module that integrates the feature representations into a common latent space containing crucial information.
    \item We carried out comprehensive experiments using the BraTS dataset for patients and the IXI dataset for healthy subjects, demonstrating the effectiveness of our approach through a detailed analysis of our designed modules.
    \item We have broadened our method to 3D by leveraging the module's adaptability and its informative feature space. This allows it to effectively utilize the fine details of 2D and the structural aspects of 3D, and it has been confirmed as a leading approach in the BraTS MRI synthesis challenge~\citep{cho2024twostageapproachbrainmr}
\end{itemize}

\section{Related Work} 
The synthesis of missing MR sequences from acquired images has recently attracted interest to date. The data-driven approach using deep learning has emerged as a promising solution to address the problem of missing data, particularly with the success of generative adversarial networks (GANs)~\citep{goodfellow2014generative}. We provide a brief overview of the image synthesis methods, classifying them into three categories: (1) one-to-one synthesis methods, (2) many-to-one synthesis methods, and (3) many-to-many unified synthesis methods. Here, we provide a technical overview of these categories to contextualize the architectural novelty of our proposed framework.

\textbf{One-to-One Synthesis Methods} synthesize the target modality from a different modality. Early work in deep learning, such as \cite{dar2019image}, developed pGAN and cGAN to generate MR images by translating between T1-weighted and T2-weighted MRI through adversarial training. \cite{xin2020multi} proposed TC-MGAN, which generates T1-weighted, post-contrast T1-weighted, and T2-FLAIR images from T2-weighted images. This method can synthesize three modalities with a single network, although the input modality remains fixed. UCD-GAN~\citep{liu2021unified} and Hyper-GAN~\citep{yang2021unified} demonstrated the possibility of a unified one-to-one synthesis approach capable of managing all modality pairs with a single network by introducing a common latent space. In our previous work, TransMI~\citep{cho2024disentangled}, we introduced an effective target modality conditioning mechanism within a unified disentanglement framework. However, a main limitation is its dependence on a single input, even though other modalities are accessible. To overcome this problem, we improve our previous approach to handle all missing scenarios through effective feature extraction and feature fusion.

\textbf{Many-to-One Synthesis Methods} integrate the information from multiple modalities to synthesize the desired target modality. These approaches primarily explored different feature fusion strategies. \cite{chartsias2017multimodal} demonstrated the shared latent space fusion approach for synthesizing the target modality from individual latent representations. HI-NET~\citep{zhou2020hi} demonstrated the effectiveness of feature fusion through the interaction of multi-level representations. DiamondGAN~\citep{li2019diamondgan} and CollaGAN~\citep{lee2020assessing} introduced feature fusion methods from individually extracted features to handle specific missing cases using a unified network.
A key development in this area was MustGAN~\citep{yurt2021mustgan}, which explicitly tried to disentangle shared and unique information. It employed a joint many-to-one stream to capture what it defined as shared information (e.g., common anatomy) and multiple parallel one-to-one streams for complementary information. However, this approach leads to an inefficient allocation of architectural complexity. The model assigns its structurally powerful many-to-one stream to the arguably simpler task of learning common features. Consequently, the less powerful one-to-one streams are delegated the more arduous task of understanding complex complementary features. Such features are generated by the non-linear interactions among sequences—a process that fundamentally challenges single-input architectures.
%Additionally, all these methods require an extra network to handle the missing scenarios. 

\begin{figure*}[!t]
\begin{center}
\includegraphics[width=1.0\textwidth]{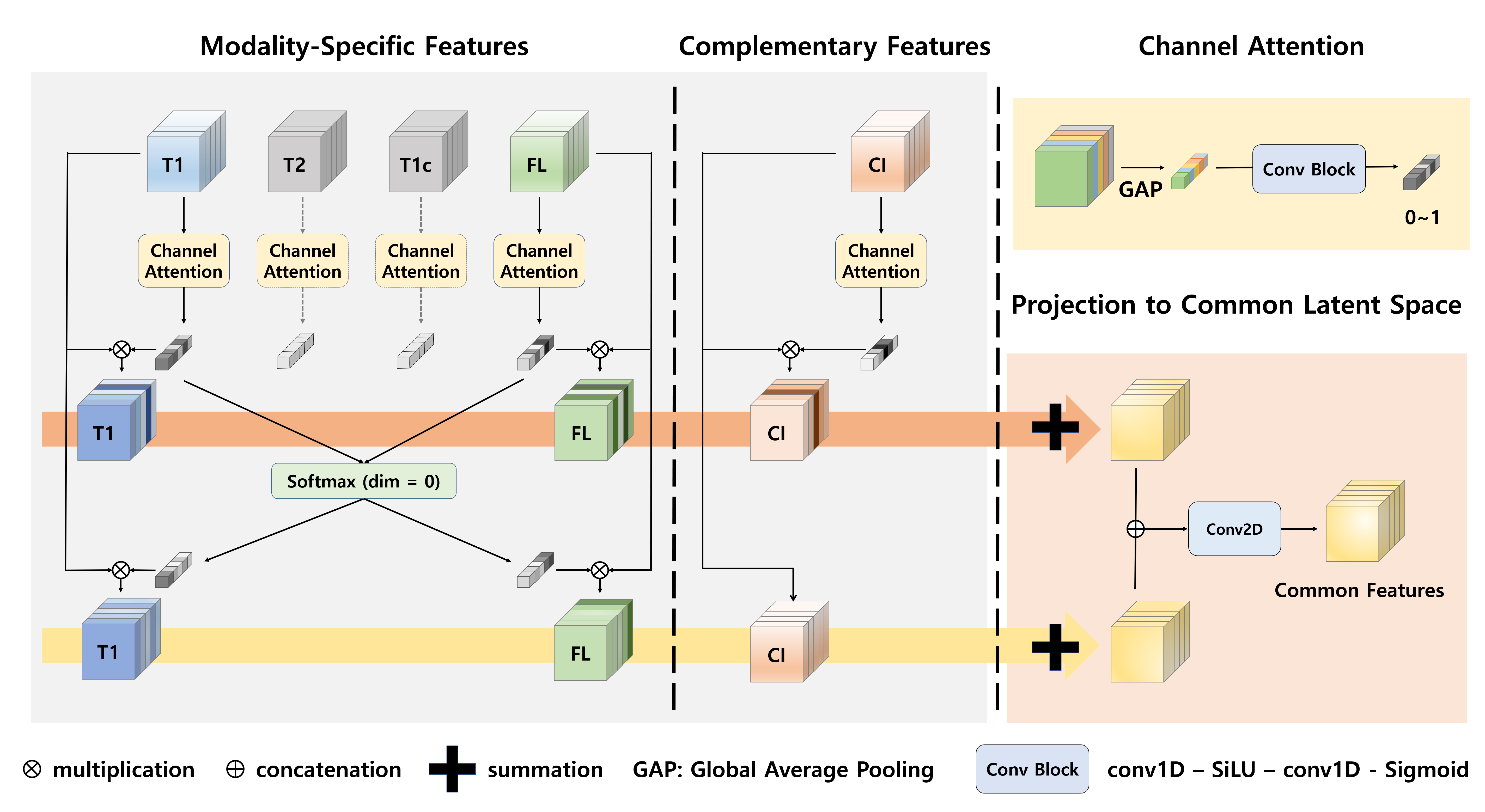}
\end{center} 
\caption{An example of the channel attention-based feature fusion module $CAFF$ for four MR sequences, which are T1, T2, T1c, and FL. $CAFF$ has two main fusion paths. (orange arrow) The first path involves the integration of significant features to emphasize essential channels, with importance maps for each feature being computed by channel attention modules. (yellow arrow) The second path functions as a residual path, applying weights to the available MR sequences using recomputed importance maps derived from the softmax operation.} 
\label{fig:caff}
\end{figure*} 

\textbf{Many-to-Many Unified Synthesis Methods} focus on handling all missing MR scenarios using a single network. Early work, such as MMGAN~\citep{sharma2019missing}, utilized a unified FCN-based architecture. Subsequent research has explored more advanced backbones, including transformers like ResViT~\citep{dalmaz2022resvit} and MM-Trans~\citep{liu2023one} for modeling long-range dependencies, and graph-based approaches like Hyper-GAE~\citep{yang2023learning} for dynamically learning inter-modal relationships. More recently, the field has adopted powerful generative models such as diffusion, which is SelfRDB~\citep{arslan2024selfconsistent}, and state-space models, which is I2I-Mamba~\citep{atli2024i2imamba}. Despite their sophistication, these methods typically rely on a single, shared feature encoding pathway, which does not architecturally enforce the explicit separation of feature types from the outset.

In contrast to all these prior works, our HF-GAN introduces a novel hybrid-fusion architecture explicitly designed for feature disentanglement. Critically, we invert the logic of earlier disentanglement efforts like mustGAN. We dedicate the many-to-one stream, the component best suited for learning inter-modal relationships, to the explicit extraction of complex complementary features. Concurrently, our parallel one-to-one streams are responsible for extracting the basic modality-specific features. This design ensures that the model's architecture is directly aligned with the nature of the information being processed, providing a more robust and effective foundation for synthesizing high-fidelity MR images, especially in challenging pathological cases.

\section{Method}

Our unified framework comprises an image generator $G$ to synthesize missing MR sequences and a discriminator $D$ to perform adversarial learning. To achieve the synthesis of all the modality compositions using a single network, our image generator consists of three key modules, including (1) hybrid-fusion encoder $Enc^{HF}$, which extracts complementary information and modality-specific information through the hybrid-fusion approach~\citep{cho2023HFTrans}, (2) channel attention-based feature fusion module, $CAFF$, which projects the extracted features into a common latent space, and (3) transformer-based modality infuser, $MI$, which facilitates the synthesis of the target modality as in our previous work~\citep{cho2024disentangled}. An overview of our framework is shown in~\autoref{fig:overview}.

\subsection{hybrid-fusion Encoder}

The first step in synthesizing missing MR sequences is to extract valuable features from available MR images. The effectiveness of this process largely depends on the encoder architecture. To facilitate the disentanglement of features, we employ a hybrid-fusion encoder $Enc^{HF}$ designed to separate feature representations according to their distinct roles. Given a set of $N$ MR sequences $I=\{I_i\}_{i=1}^N$ and a binary availability mask $AS=\{AS_i\in \{0,1\}\}_{i=1}^N$, the input to our framework is a set of masked images $X=\{X_i\}_{i=1}^N=\{I_i\cdot AS_i\}_{i=1}^N$. The hybrid-fusion encoder consists of two distinct pathways: \\

\noindent \textbf{Modality-Specific Pathway:} $N$ parallel 1-channel CNN encoders $Enc^{M}_i (i=1,...,N)$ extract modality-specific features $F_i$ from each available MR sequence individually. This pathway focuses on capturing the unique structural information and contrast characteristics of each modality:
\begin{equation}
    F_i = Enc^{M}_i(X_i) \quad \forall i \ \text{where} \ AS_i=1.
\end{equation}

\noindent \textbf{Complementary Pathway:} A single early fusion (multi-channel) CNN encoder $Enc^{C}$  extracts complementary features $F_0$ by processing all available sequences simultaneously. This pathway is activated only when multiple sequences are available $(\Sigma AS>1)$ and is designed to learn the complex inter-dependencies and relationships between modalities, which is especially critical for identifying pathological regions. The input is a channel-wise concatenation of the available masked images, denoted as $X_{cat}$:

\begin{equation}
    F_0 = 
    \begin{cases} 
        \text{Enc}^{C}(X_{\text{cat}}) & \text{if } \Sigma AS > 1 \\
        0 & \text{if } \Sigma AS \le 1. 
    \end{cases}
\end{equation}

Both encoders ($Enc^{C}$ and $Enc^{M}_i$) share an identical architecture composed of five residual blocks, with the only difference being the number of input channels. The final output of the encoder stage is a collection of feature representations, $F=\{F_0,F_1,...,F_N\}$, where features corresponding to unavailable modalities are zeroed out.

\subsection{Channel Attention-based Feature Fusion}
The collection of features $F$ extracted from the hybrid-fusion encoder must be integrated into a single, modality-agnostic feature map in a common latent space $z$. This is challenging given the $2^N-2$ possible combinations of available sequences. To address this, we propose the Channel Attention-based Feature Fusion (CAFF) module, which dynamically highlights salient information and integrates the distinct feature types. The process, illustrated in~\autoref{fig:caff}, can be broken down into the following steps: \\ 

\noindent \textbf{Importance Map Generation:} First, for each non-zero feature map $F_i$ (for $i=0,...,N$), we compute a channel-wise importance map (attention vector) $M_i$ using a a channel attention (CA) block. The CA block consists of Global Average Pooling (GAP) followed by a small multi-layer perceptron (MLP) composed of 1D convolutions and a sigmoid activation $\sigma$:
\begin{equation}
    M_i = \text{CA}(F_i) = \sigma(\text{Conv1D}(\delta(\text{Conv1D}(\text{GAP}(F_i))))),
\end{equation}
where $\delta$ is the SiLU activation function. \\

\noindent \textbf{Primary Feature Integration:} The first fusion path (orange arrow in~\autoref{fig:caff}) integrates the most significant features by weighting each feature map with its corresponding importance map. This path emphasizes essential channels in both the complementary and modality-specific features. The result $z_{primary}$ is formed by the element-wise product ($\otimes$) and summation:
\begin{equation}
    z_{primary} = (F_0\otimes M_0) + \Sigma^{N}_1(F_i\otimes M_i).
\end{equation}
Note that unavailable sequences have $F_i=0$, so they do not contribute to the sum. \\

\noindent \textbf{Residual Feature Integration:} The second path (yellow arrow in~\autoref{fig:caff}) serves as a residual connection to preserve robust structural information, ensuring that a stable baseline of anatomical features is maintained even if the primary attention path incorrectly down-weights them. We compute a re-weighted importance map $M_{i}^{'}$, derived by applying a softmax operation across the attention maps of the available modality-specific features. This adjusts their relative significance. The complementary features $F_0$ are passed through directly:
\begin{equation}
    M_{i}^{'}={{exp(M_i)} \over {\Sigma^{N}_{j=1,AS_j=1}exp(M_i)}},
\end{equation}

\begin{equation}
    z_{residual} = F_0 + \Sigma^{N}_1(F_i\otimes M_{i}^{'}).
\end{equation}

\noindent \textbf{Final Projection:} Finally, the features from both pathways are combined to form the common latent space representation $z$. The primary features $z_{primary}$, which contain the dynamically selected information, are concatenated with the residual features $z_{residual}$, which provide a stable structural baseline. This combined feature map is then passed through a final 2D convolutional layer $Conv_{proj}$, which serves as a projection head. This layer reduces the channel dimensionality and learns the optimal integration of the two feature streams:
\begin{equation}
    z = Conv_{proj}(Concat(z_{primary}, z_{residual})).
\end{equation}
This two-pathway fusion within CAFF ensures that crucial features are capitalized upon while preventing the loss of important channel information, allowing for a robust projection into a consistent space regardless of the input composition.

\begin{figure}[!t]
\begin{center}
\includegraphics[width=1.0\columnwidth]{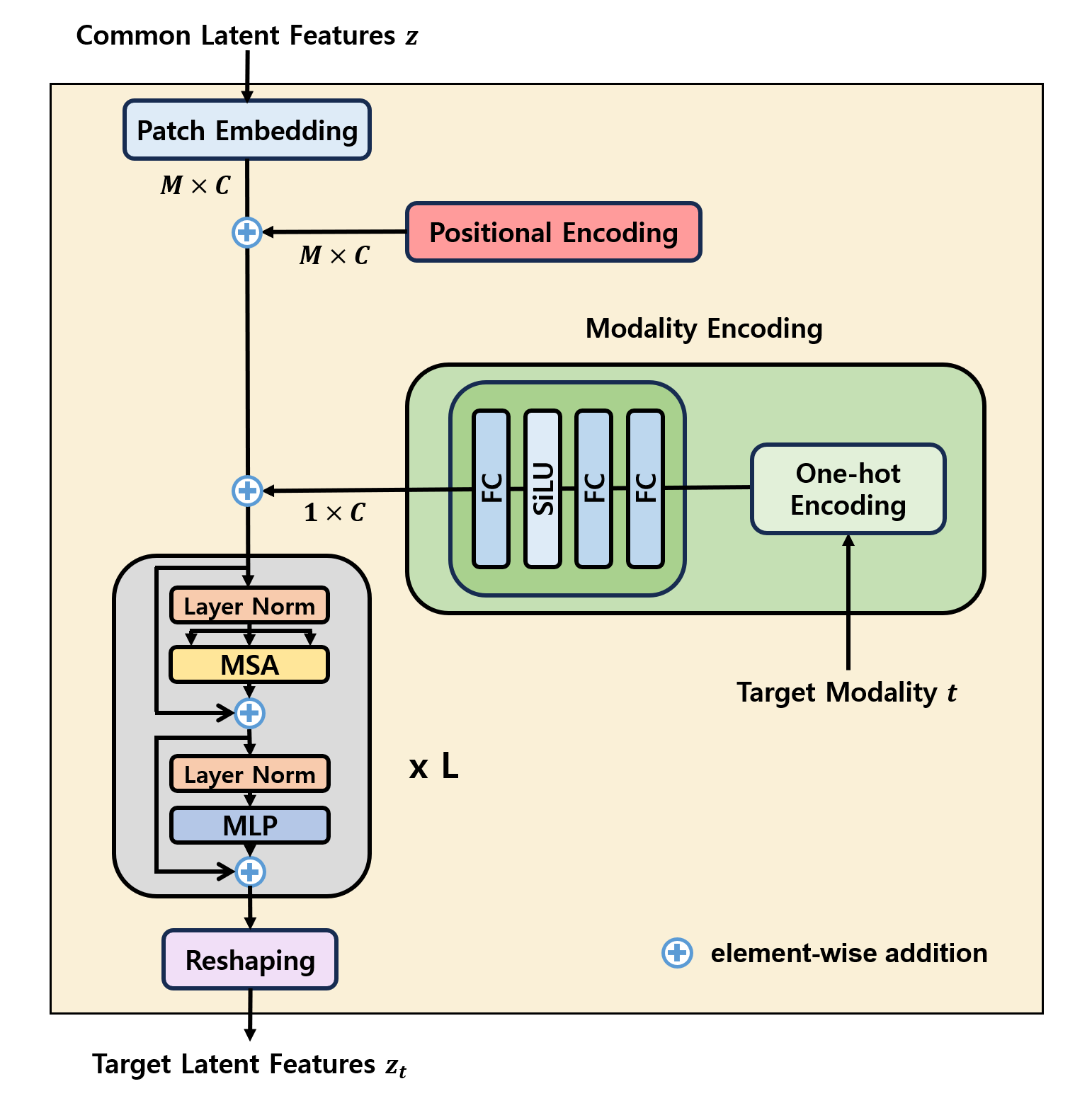}
\end{center} 
\caption{The structure of the modality infuser $MI$. The target modality is encoded by one-hot encoding.} 
\label{fig:mc}
\end{figure}

\begin{figure*}[!t]
\begin{center}
\includegraphics[width=1.0\textwidth]{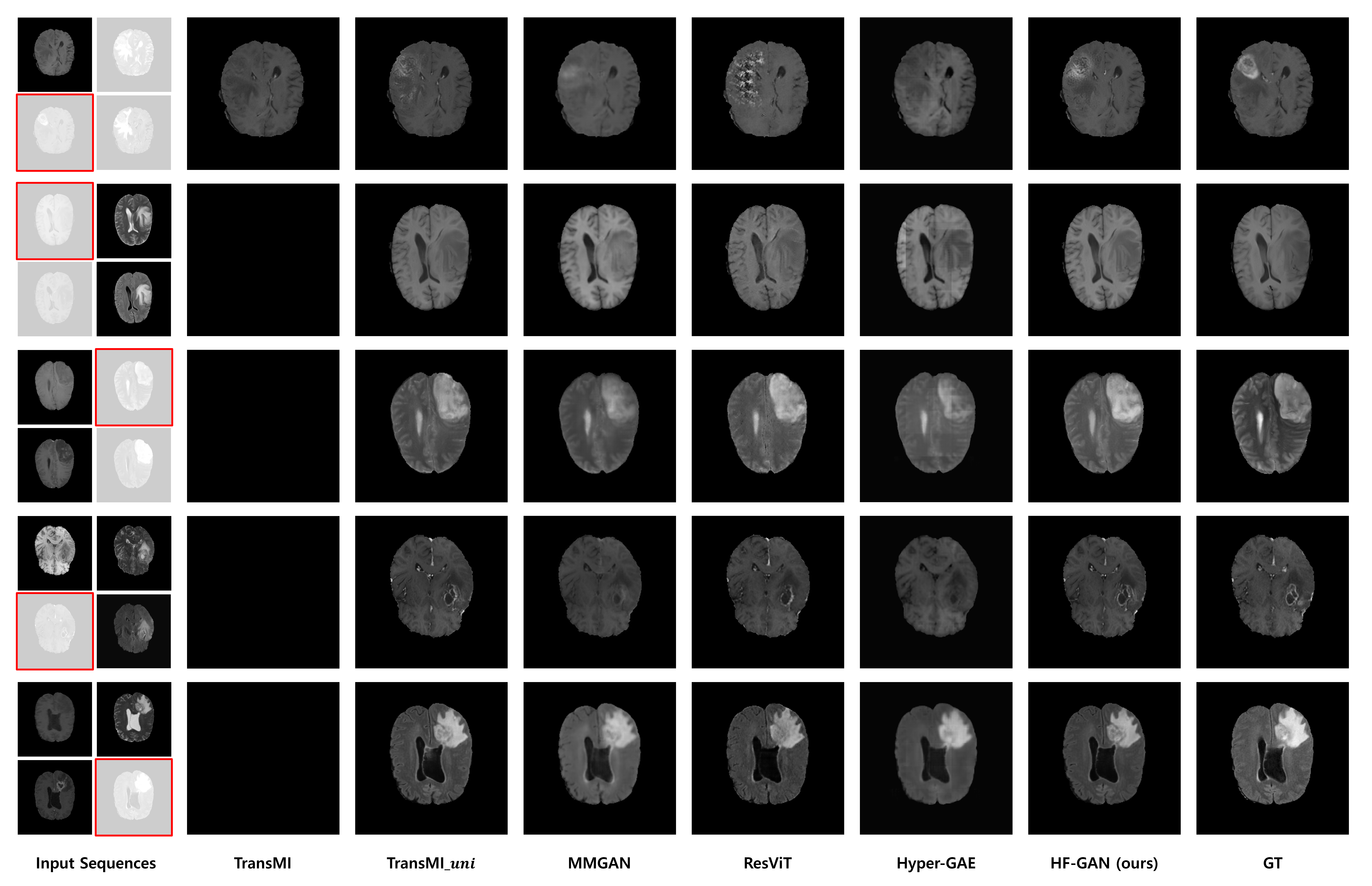}
\end{center} 
\caption{An example of synthesized results for the missing MR sequence on the BraTS dataset. The first column represents the input MR sequences: T1, T2, FL, and T1c are located from the top-left in a clockwise direction. The target MR sequence (GT) is highlighted by a red border.  } 
\label{fig:qual_brats}
\end{figure*} 

\subsection{Modality Infuser}

To synthesize the target missing MR sequence $t \in \{1,...,N\}$, auxiliary information about the target is necessary. Instead of simply adding target indicators as extra channels~\citep{liu2021unified, mirza2014conditional}, which can degrade performance, we employ a powerful conditioning approach called the Modality Infuser (MI). The $MI$ transforms the feature representations from the common latent space $z$ to a target-specific latent space $z_t$ through a series of learnable, attention-based operations. The architecture is detailed in~\autoref{fig:mc} and can be formalized as follows: \\

\noindent \textbf{Tokenization and Encoding}: The common latent feature map $z \in \mathbb{R}^{h\times w\times c}$ is first flattened and projected into $M$ 1D tokens $f_0 \in \mathbb{R}^{M\times C}$ via a patch embedding layer $Emb$:
\begin{equation}
    f_0 = Emb(z).
\end{equation}
To provide the model with spatial information, a learnable positional encoding $PE\in \mathbb{R}^{M\times C}$ is added to the tokens. Concurrently, the target modality index $t$ is converted into a one-hot vector $v_t \in \mathbb{R}^N$, where N is the total number of modalities. This vector is then processed by a multi-layer perceptron (MLP) to generate the final modality encoding vector $ME\in \mathbb{R}^{1\times C}$:
\begin{equation}
    ME = \text{MLP}(v_t).
\end{equation}
The modality encoding is then broadcast and added to all tokens, effectively shifting the entire feature set towards the target modality's latent subspace. The input to the transformer stack $s_{0}$ is thus:
\begin{equation}
    s_{0} = f_0 + PE + ME.
\end{equation}
\noindent \textbf{Transformer Blocks:} The encoded tokens $s_{0}$ are processed through $L$(=4) successive transformer blocks. Each block $l$ (from $l=1$ to $L$) consists of a Multi-Head Self-Attention (MSA) module and a feed-forward MLP, both with residual connections and Layer Normalization (LN):
\begin{equation}
\begin{split}
    &s_l^{'}=\text{MSA}(\text{LN}(s_{l-1}))+s_{l-1}, \\
    &s_l=\text{MLP}(\text{LN}(s_{l}^{'}))+s_{l}^{'}.
\end{split}
\end{equation}
This deep, attention-based processing refines the tokens, ensuring they are contextually consistent with the target modality. \\

\noindent \textbf{Reshaping and Output:} After the final transformer block, the processed tokens $s_L$ are reshaped back into the original spatial dimensions to form the final target latent features $z_t$. These target-specific features are then fed into the CNN decoder $Dec$, which has an architecture symmetric to the encoders, to synthesize the final target image $\tilde{y}_t$:
\begin{equation}
    \tilde{y}_t = Dec(z_t).
\end{equation}

%%%%%%%%%%%%%%%%%%%%%%%%%%%%%
\begin{table*}[t]
\caption{\label{table:quan_brats} Quantitative evaluation results for multisequence MRI synthesis on the BraTS dataset. An asterisk (*) indicates a statistically significant difference (Wilcoxon signed-rank test, p$<$0.05).}
\begin{center}
\resizebox{1.0\textwidth}{!}{
\begin{tabular}{>{\centering}p{5mm}>{\centering}p{5mm}>{\centering}p{5mm}>{\centering}p{5mm}|>{\centering}p{19mm}>{\centering}p{25mm}|>{\centering}p{19mm}>{\centering}p{25mm}|>{\centering}p{19mm}>{\centering}p{25mm}|>{\centering}p{19mm}>{\centering}p{25mm}|>{\centering}p{19mm}>{\centering}p{25mm}|>{\centering}p{19mm}>{\centering\arraybackslash}p{25mm}}
\hline

\multicolumn{4}{c|}{Input sequences} & \multicolumn{2}{c|}{TransMI} & \multicolumn{2}{c|}{$\textrm{TransMI}_{uni}$}& \multicolumn{2}{c|}{MMGAN} & \multicolumn{2}{c|}{ResViT} & \multicolumn{2}{c|}{Hyper-GAE} & \multicolumn{2}{c}{HF-GAN (ours)} \\ \cline{1-16} 
 T1 & T2 & T1c & FL & PSNR & SSIM & PSNR & SSIM & PSNR & SSIM & PSNR & SSIM & PSNR & SSIM & PSNR & SSIM \\
\hline
\hline

\checkmark & - & - & - & 29.30 (4.10)* & 0.9256 (0.0405)*& 29.81 (4.06)* & 0.9316 (0.0374)*& 29.30 (4.30)* & 0.9109 (0.0394)*& 28.84 (3.90)* & 0.9137 (0.0441)*& 29.48 (3.86)* & 0.9279 (0.0352)*& \textbf{30.00 (4.07)} & \textbf{0.9324 (0.0373)} \\
- & \checkmark - & - & - & 29.61 (4.20)* & 0.9329 (0.0349)*& 29.53 (4.40)* & 0.9337 (0.0350)*& 28.81 (4.65)* & 0.9136 (0.0416)*& 28.42 (4.15)* & 0.9135 (0.0420)*& 29.17 (4.15)* & 0.9333 (0.0334)*& \textbf{29.92 (4.33)} & \textbf{0.9372 (0.0330)} \\
- & - & \checkmark & - & 28.73 (4.22)* & 0.9228 (0.0426)*& 29.09 (4.12)* & 0.9280 (0.0388)*& 28.47 (4.32)* & 0.9244 (0.0383)*& 28.10 (3.86)* & 0.9116 (0.0425)*& 28.63 (3.84)* & 0.9252 (0.0358)*& \textbf{29.28 (4.13)} & \textbf{0.9295 (0.0381)} \\
- & - & - & \checkmark & 28.67 (4.14)* & 0.9188 (0.0440)*& 29.40 (4.05)* & 0.9288 (0.0361)*& 28.70 (4.15)* & 0.9070 (0.0420)*& 28.32 (3.83)* & 0.9114 (0.0424)*& 28.86 (3.89)* & 0.9274 (0.0361)*& \textbf{29.53 (4.03)} & \textbf{0.9297 (0.0353)} \\ \hline
\multicolumn{4}{c|}{Average} & 29.08 (4.19)* & 0.9250 (0.0410)*& 29.45 (4.17)* & 0.9305 (0.0369)*& 28.82 (4.37)* & 0.9140 (0.0409)*& 28.42 (3.95)* & 0.9125 (0.0428)*& 29.04 (3.95)* & 0.9285 (0.0353)*& \textbf{29.68 (4.15)} & \textbf{0.9322 (0.0361)} \\ \hline
\checkmark & \checkmark & - & - & - & - & 30.57 (4.37)* & 0.9358 (0.0331)*& 29.88 (4.63)* & 0.9107 (0.0355)*& 29.72 (4.16)* & 0.9203 (0.0387)*& 30.48 (4.08)* & \textbf{0.9391 (0.0303)}& \textbf{31.07 (4.34)} & 0.9387 (0.0316) \\
\checkmark & - & \checkmark & - & - & - & 29.39 (3.81)* & 0.9302 (0.0366)*& 28.83 (4.08)* & 0.9260 (0.0361)*& 28.61 (3.62)* & 0.9158 (0.0425)*& 29.24 (3.62)* & 0.9325 (0.0330)*& \textbf{29.80 (3.88)} & \textbf{0.9329 (0.0355)} \\
\checkmark & - & - & \checkmark & - & - & 30.82 (3.74)* & 0.9454 (0.0281)*& 30.30 (3.73)* & 0.9179 (0.0343)*& 30.00 (3.58)* & 0.9346 (0.0319)*& 30.61 (3.51)* & 0.9464 (0.0256)*& \textbf{31.22 (3.72)} & \textbf{0.9471 (0.0275)} \\
- & \checkmark & \checkmark & - & - & - & 29.60 (4.58)* & 0.9388 (0.0350)*& 28.78 (4.90)* & 0.9342 (0.0370)*& 28.80 (4.38)* & 0.9248 (0.0397)*& 29.38 (4.23)* & 0.9391 (0.0322)*& \textbf{30.11 (4.53)} & \textbf{0.9429 (0.0327)} \\
- & \checkmark & - & \checkmark & - & - & 30.27 (4.57)* & 0.9446 (0.0298)*& 29.44 (4.69)* & 0.9180 (0.0405)*& 29.46 (4.32)* & 0.9315 (0.0340)*& 29.80 (4.25)* & 0.9441 (0.0279)*& \textbf{30.75 (4.51)} & \textbf{0.9480 (0.0280)} \\
- & - & \checkmark & \checkmark & - & - & 29.93 (4.10)* & 0.9446 (0.0293)*& 29.14 (4.25)* & 0.9382 (0.0307)*& 29.16 (4.02)* & 0.9332 (0.0329)*& 29.41 (3.83)* & 0.9433 (0.0276)*& \textbf{30.38 (4.12)} & \textbf{0.9477 (0.0278)} \\ \hline
\multicolumn{4}{c|}{Average} & - & - & 30.10 (4.24)* & 0.9399 (0.0326)*& 29.39 (4.43)* & 0.9242 (0.0371)*& 29.29 (4.05)* & 0.9267 (0.0375)*& 29.82 (3.97)* & 0.9407 (0.0299)*& \textbf{30.55 (4.22)} & \textbf{0.9429 (0.0312)} \\ \hline
\checkmark & \checkmark & \checkmark & - & - & - & 29.71 (4.36)* & 0.9302 (0.0333)*& 28.76 (4.71)* & 0.9243 (0.0367)*& 28.98 (4.03)* & 0.9149 (0.0389)*& 29.62 (4.08)* & 0.9320 (0.0320)*& \textbf{30.27 (4.38)} & \textbf{0.9349 (0.0313)} \\
\checkmark & \checkmark & - & \checkmark & - & - & 31.93 (4.31)* & 0.9491 (0.0266)*& 31.16 (4.13)* & 0.9023 (0.0271)*& 31.28 (4.02)* & 0.9387 (0.0304)*& 31.55 (3.89)* & 0.9481 (0.0253)*& \textbf{32.55 (4.30)} & \textbf{0.9521 (0.0253)} \\
\checkmark & - & \checkmark & \checkmark & - & - & 30.47 (3.19)* & 0.9501 (0.0251)*& 29.81 (3.20)* & 0.9413 (0.0255)*& 29.64 (3.08)* & 0.9411 (0.0282)*& 30.17 (3.01)* & 0.9483 (0.0234)*& \textbf{31.06 (3.27)} & \textbf{0.9533 (0.0239)} \\
- & \checkmark & \checkmark & \checkmark & - & - & 30.13 (5.06)* & 0.9560 (0.0267)*& 28.96 (5.24)* & 0.9493 (0.0299)*& 29.36 (4.84)* & 0.9440 (0.0307)*& 29.45 (4.55)* & 0.9526 (0.0255)*& \textbf{30.69 (5.01)} & \textbf{0.9595 (0.0248)} \\ \hline
\multicolumn{4}{c|}{Average} & - & - & 30.56 (4.36)* & 0.9463 (0.0298)*& 29.67 (4.49)* & 0.9293 (0.0351)*& 29.81 (4.13)* & 0.9347 (0.0343)*& 30.20 (4.01)* & 0.9452 (0.0279)*& \textbf{31.14 (4.37)} & \textbf{0.9500 (0.0280)} \\ \hline\hline
\multicolumn{4}{c|}{Average (all)} & - & - & 29.89 (4.25)* & 0.9368 (0.0346)*& 29.19 (4.42)* & 0.9205 (0.0389)*& 28.99 (4.05)* & 0.9218 (0.0403)*& 29.54 (3.99)* & 0.9361 (0.0328)*& \textbf{30.27 (4.25)} & \textbf{0.9393 (0.0336)} \\

\hline
\end{tabular}
}
\end{center}

\end{table*}
%%%%%%%%%%%%%%%%%%%%%%%%%%%%%

\subsection{Training loss}
We assume spatially co-registered MR sequences, so that the synthesized images can be directly compared with the actual image $x_t$ using a L1 loss defined as
\begin{equation}
    L_{rec} = | x_t - \tilde{y}_t |.
\end{equation}
The reconstruction loss ensures that the feature representation comprehensively captures the characteristics of the training data across MR sequences. The synthesized images $\tilde{y}_t$ are used again to enhance the preservation of shape structures through a cycle-consistency loss $L_{cyc}$~\citep{zhu2017unpaired} defined as 
\begin{equation}
    L_{cyc} = | x_c - G(\hat{X}_t,\hat{AS},c)) |,
\end{equation}
where $c$ denotes one of the available MR sequences utilized to synthesize $\tilde{y}_t=G(X,AS,t)$ ($x_c \in X$ and $c \in AS$). Meanwhile, $\hat{X}_t$ and $\hat{AS}$ indicate the newly assembled MR images and sequences, which incorporate the synthesized target sequence ($\tilde{y}_t \in \hat{X_t}$ and $t \in \hat{AS}$) while omitting the new target sequence ($c \notin \hat{AS}$).

Furthermore, the feature similarity loss $L_{sim}$ is applied to establish a common latent space by promoting alignment within the latent space for different input composition of the same subject, and it is defined as

\begin{equation}
    L_{sim} = \frac{z \cdot \hat{z}}{\| z\| \cdot \| \hat{z} \|},
\end{equation}
where $\hat{z}$ denotes the latent feature representations extracted from the images of the same subject as $z$, but with a different input composition $\hat{AS}$.

For the adversarial training process, we use the discriminator $D$ of PatchGAN~\citep{isola2017image}. This discriminator is tasked with discriminating between real and synthetic images and also classifies their modalities. The adversarial loss and the classification loss are defined as follows:
\begin{equation}
    % L_{adv}= L_{BCE}(D(\tilde{y}_t),1)
    L_{adv}= \mathbb{E}_{x}[logD(x)] + \mathbb{E}_{\tilde{y}}[1-logD(\tilde{y})],
\end{equation}
\begin{equation}
    % L_{aux}=L_{CE}(D_{aux}(x_c),c) + L_{CE}(D_{aux}(\tilde{y}_t),t)
    L_{cls}=\mathbb{E}_{x,c}[-logD_{cls}(c|x)] + \mathbb{E}_{\tilde{y},t}[-logD_{cls}(t|\tilde{y})].
\end{equation}
% where $L_{BCE}$ and $L_{CE}$ represent binary cross entropy loss and cross entropy loss respectively.
Incorporating an auxiliary modality classification task allows training with a single discriminator and helps to construct probability distributions between different modalities~\citep{odena2017conditional}.

The total losses of the image generator and discriminator are defined as the weighted sum of the loss components:
\begin{equation}
    L_{G}=\alpha L_{rec}+\beta L_{sim}+\gamma L_{cyc}+\lambda_1 L_{adv}+\lambda_2 L_{cls},
\end{equation}

\begin{equation}
    % L_{D}=\lambda_3 \cdot (L_{BCE}(D(x),1) + L_{BCE}(D(\tilde{y}_t),0)) + \lambda_4 \cdot L_{aux}
    L_{D}=-\lambda_3 \cdot L_{adv} + \lambda_4 \cdot L_{cls},
\end{equation}
where $\alpha$, $\beta$, $\gamma$, $\lambda_1$, $\lambda_2$, $\lambda_3$, and $\lambda_4$ are weightings for each loss item.

\subsection{Training scheme}
For stable and robust training of our framework, we carefully select available sequences ${AS}$ for $L_{rec}$ and available sequences with synthesized image $\hat{AS}$ for $L_{sim}$ and $L_{cyc}$. Half of the mini-batch is used exclusively for the most difficult scenarios to compensate for the absent complementary information ($\Sigma AS = 1$ and $\Sigma \hat{AS} = 1$). The other half of the mini-batch randomly selects the initial available MR sequences except for extreme conditions ($\Sigma AS > 1$). To improve the extraction of complementary information and to construct useful common spaces, we select scenarios that produce the most informative feature representations ($\hat{AS} = N-1$). Furthermore, for each subject, the network parameters are updated $N$ times by employing all possible target sequences $t$, maximizing the use of MR composition diversity. This curriculum-like sampling strategy was implemented to ensure the model develops robust representations from information-scarce inputs and prevents it from overfitting to easier, multi-sequence cases.

%%%%%%%%%%%%%%%%%%%%%%%%%%%%%
\begin{table*}[t]
\caption{\label{table:quan_ixi} Quantitative evaluation results for multisequence MRI synthesis on the IXI dataset. An asterisk (*) indicates a statistically significant difference (Wilcoxon signed-rank test, p$<$0.05).}
\begin{center}
\resizebox{1.0\textwidth}{!}{
\begin{tabular}{>{\centering}p{5mm}>{\centering}p{5mm}>{\centering}p{5mm}|>{\centering}p{19mm}>{\centering}p{25mm}|>{\centering}p{19mm}>{\centering}p{25mm}|>{\centering}p{19mm}>{\centering}p{25mm}|>{\centering}p{19mm}>{\centering}p{25mm}|>{\centering}p{19mm}>{\centering}p{25mm}|>{\centering}p{19mm}>{\centering\arraybackslash}p{25mm}}
\hline

\multicolumn{3}{c|}{Input sequences} & \multicolumn{2}{c|}{TransMI} & \multicolumn{2}{c|}{$\textrm{TransMI}_{uni}$}& \multicolumn{2}{c|}{MMGAN} & \multicolumn{2}{c|}{ResViT} & \multicolumn{2}{c|}{Hyper-GAE} & \multicolumn{2}{c}{HF-GAN (ours)} \\ \cline{1-15} 
 T1 & T2 & PD & PSNR & SSIM & PSNR & SSIM & PSNR & SSIM & PSNR & SSIM & PSNR & SSIM & PSNR & SSIM \\
\hline \hline

\checkmark & - & - & 25.38 (2.67)* & 0.8555 (0.0609)*& 25.34 (2.63)* & 0.8574 (0.0608)*& 24.11 (2.13)* & 0.8325 (0.0612)*& 25.16 (2.60)* & 0.8502 (0.0597)*& 25.15 (2.42)* & 0.8556 (0.0593)*& \textbf{25.47 (2.68)} & \textbf{0.8575 (0.0605)} \\
- & \checkmark & - & 29.20 (3.63)* & 0.9221 (0.0415)*& 29.18 (3.69)* & 0.9222 (0.0415)* & 27.87 (3.37)* & 0.9064 (0.0468)*& 28.89 (3.73)* & \textbf{0.9249 (0.0424)}& 28.57 (3.14)* & 0.9215 (0.0397)*& \textbf{29.36 (3.81)} & 0.9230 (0.0420) \\
- & - & \checkmark & 28.71 (3.02)* & 0.9179 (0.0392)*& 28.70 (3.08)* & 0.9176 (0.0393)*& 27.32 (2.67)* & 0.9005 (0.0431)*& 28.39 (3.11)* & 0.9181 (0.0396)*& 28.25 (2.56)* & \textbf{0.9199 (0.0363)*}& \textbf{28.87 (3.16)} & 0.9177 (0.0397) \\ \hline
\multicolumn{3}{c|}{Average} & 27.76 (3.56)* & 0.8985 (0.0570)*& 27.74 (3.59)* & 0.8990 (0.0565)*& 26.43 (3.23)* & 0.8798 (0.0610)*& 27.48 (3.58)* & 0.8978 (0.0587)*& 27.32 (3.13)* & 0.8990 (0.0555)& \textbf{27.90 (3.68)} & \textbf{0.8994 (0.0567)} \\ \hline
\checkmark & \checkmark & - & - & -& 32.64 (2.75)* & 0.9463 (0.0230)*& 30.85 (2.55)* & 0.9396 (0.0249)*& 31.98 (2.72)* & 0.9481 (0.0219)*& 31.24 (2.59)* & 0.9426 (0.0232)*& \textbf{33.09 (2.86)} & \textbf{0.9487 (0.0225)} \\
\checkmark & - & \checkmark & - & -& 31.58 (2.47)* & 0.9343 (0.0261)*& 30.12 (2.27)* & 0.9274 (0.0268)*& 30.91 (2.45)* & 0.9356 (0.0245)*& 30.37 (2.18)* & 0.9348 (0.0246)*& \textbf{31.96 (2.53)} & \textbf{0.9358 (0.0257)} \\
- & \checkmark & \checkmark & - & - & 27.38 (2.64)* & 0.9163 (0.0430)*& 26.47 (2.55)* & 0.8957 (0.0470)*& 26.84 (2.68)* & 0.9118 (0.0449)*& \textbf{27.60 (2.61)*} & 0.9169 (0.0408)*& 27.53 (2.77) & \textbf{0.9179 (0.0431)} \\ \hline
\multicolumn{3}{c|}{Average} & - & -& 30.53 (3.47)* & 0.9323 (0.0342)*& 29.14 (3.12)* & 0.9209 (0.0390)*& 29.91 (3.43)* & 0.9318 (0.0355)*& 29.74 (2.91)* & 0.9314 (0.0324)*& \textbf{30.86 (3.63)} & \textbf{0.9341 (0.0342)} \\ \hline\hline
\multicolumn{3}{c|}{Average (all)} & - & -& 28.67 (3.79)* & 0.9101 (0.0526)*& 27.34 (3.44)* & 0.8935 (0.0580)*& 28.29 (3.71)* & 0.9091 (0.0545)*& 28.13 (3.27)* & 0.9098 (0.0513)*& \textbf{28.89 (3.92)} & \textbf{0.9110 (0.0529)} \\

\hline
\end{tabular}
}
\end{center}

\end{table*}
%%%%%%%%%%%%%%%%%%%%%%%%%%%%%

\begin{figure*}[!t]
\begin{center}
\includegraphics[width=1.0\textwidth]{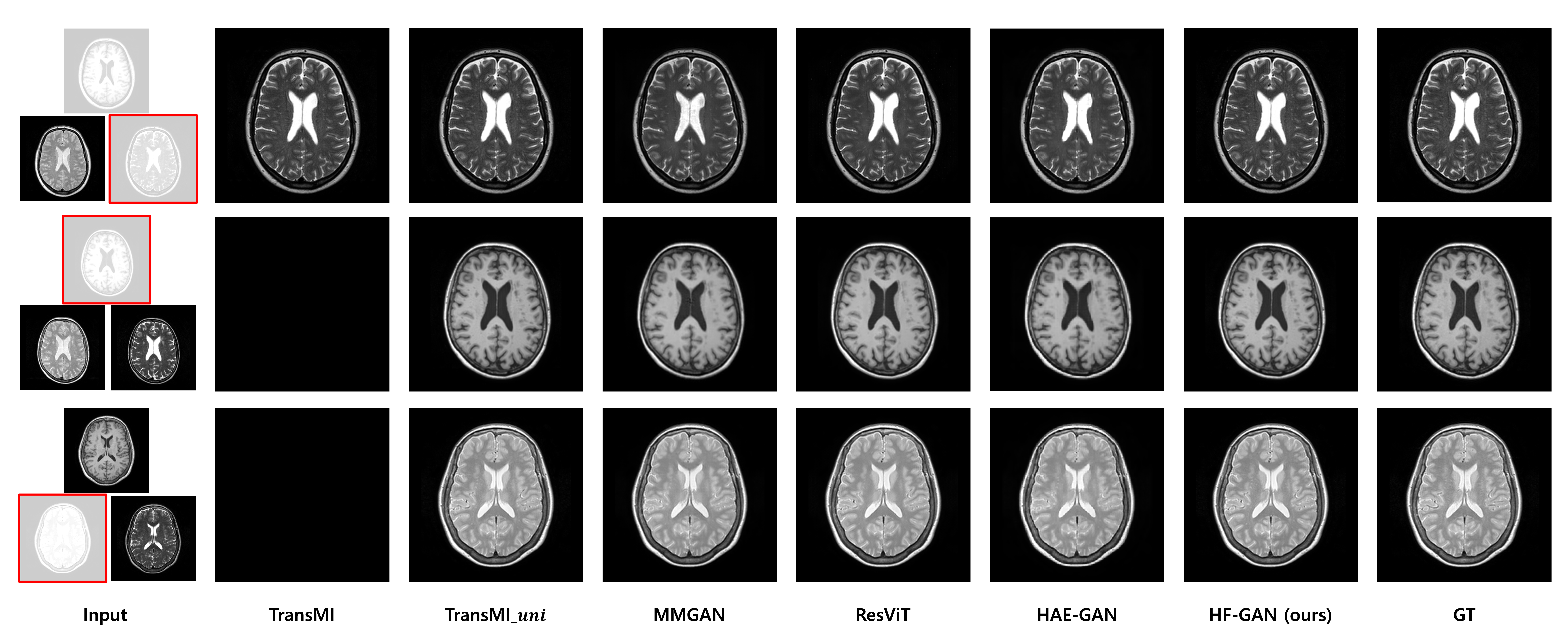}
\end{center} 
\caption{An example of synthesized results for the missing MR sequence on the IXI dataset. The first column represents the input MR sequences: T1, T2, and PD are located from the top in a clockwise direction. The target MR sequence (GT) is highlighted by a red border.} 
\label{fig:qual_ixi}
\end{figure*} 

\subsection{Implementation}
Our framework is implemented using PyTorch, and all experiments in this study are carried out on an NVIDIA RTX 3090 GPU (24GB VRAM). The networks are optimized in 100 epochs using an Adam optimizer~\citep{kingma2014adam} with $\beta_1=0.9$, $\beta_2=0.999$, and a batch size of 24 using gradient accumulation and half precision (fp16). We set initial learning rates of 0.0001 and 0.00001 for the generator and discriminator, respectively, with the cosine annealing scheduler~\citep{loshchilov2016sgdr}, which has a short warm-up phase and then decays to zero afterward. The weighting of the loss components is configured as follows: $\alpha = 10$, $\beta = 1$, $\gamma = 1$, $\lambda_1 = 0.25$, $\lambda_2 = 0.25$, $\lambda_3 = 0.25$, and $\lambda_4 = 0.25$. Our source code is available on \href{https://github.com/SSTDV-Project/HF-GAN}{https://github.com/SSTDV-Project/HF-GAN}.

\section{Experiments}
\subsection{Datasets}
To validate the effectiveness of the proposed method, we have experimented with two public multisequence MRI datasets, the BraTS 2018 dataset~\citep{bakas2017advancing,bakas2018identifying,menze2014multimodal} and the IXI dataset~\footnote{https://brain-development.org/ixi-dataset}.

\textbf{BraTS 2018} is a patient brain MRI dataset comprising multiple institutional MRI scans of glioblastoma and lower-grade glioma. It consists of 285 subjects, each with four MR sequences: T1-weighted (T1), T2-weighted (T2), post-contrast T1-weighted (T1c), and T2-FLAIR (FL) MRIs. We partitioned the dataset into three subsets: 180 subjects for training, 20 subjects for validation, and 85 subjects for testing. For further data imputation experiments, the training dataset is divided into 80 subjects for training our framework and 100 subjects for training the segmentation network. Intensity normalization is performed by linearly scaling of the original intensities to the range $[-1, 1]$. During training, 2D axial slices are used, excluding those with less than 2000 brain pixels. The size of the final images is set to $240 \times 240$. The number of slices used for the training, validation, and test sets is 22,801, 2,511, and 10,745, respectively. 

%%%%%%%%%%%%%%%%%%%%%%%%%%%%%
\begin{table*}[t]
\caption{\label{table:seg} Brain tumor segmentation results with data imputation on the BraTS dataset. The results represent the average Dice score over 85 test subjects and WT, TC, and ET stand for whole tumor, tumor core, and enhanced tumor, respectively. An asterisk (*) indicates a statistically significant difference (Wilcoxon signed-rank test, p$<$0.05).}
\begin{center}
\resizebox{1.0\textwidth}{!}{
\begin{tabular}{>{\centering}p{5mm}>{\centering}p{5mm}>{\centering}p{5mm}>{\centering}p{5mm}|>{\centering}p{15mm}>{\centering}p{15mm}>{\centering}p{15mm}|>{\centering}p{15mm}>{\centering}p{15mm}>{\centering}p{15mm}|>{\centering}p{15mm}>{\centering}p{15mm}>{\centering}p{15mm}|>{\centering}p{15mm}>{\centering}p{15mm}>{\centering}p{15mm}|>{\centering}p{15mm}>{\centering}p{15mm}>{\centering\arraybackslash}p{15mm}}
\hline

\multicolumn{4}{c|}{\multirow{2}{*}{Available sequences}} & \multicolumn{15}{c}{Data imputation method} \\ \cline{5-19}
& & & & \multicolumn{3}{c|}{without Synthesis} & \multicolumn{3}{c|}{Synthesized from MMGAN} & \multicolumn{3}{c|}{Synthesized from ResViT} & \multicolumn{3}{c|}{$\textrm{Synthesized from Hyper-GAE}$}& \multicolumn{3}{c}{Synthesized from HF-GAN (ours)} \\ \hline
 T1 & T2 & T1c & FL & WT & TC & ET & WT & TC & ET & WT & TC & ET & WT & TC & ET & WT & TC & ET \\
\hline
\hline

\checkmark & - & - & - & 0.0000* (0.0000) & 0.0000* (0.0000) & 0.0824* (0.2749)& 0.5323* (0.2099) & 0.3917* (0.1761) & 0.1212* (0.2754)& 0.1402* (0.2057) & 0.0699* (0.1005) & 0.0847* (0.2745)& \textbf{0.6283* (0.2071)} & \textbf{0.4756* (0.1950)} & 0.0922* (0.2752)& 0.5806 (0.2281) & 0.4335 (0.2112) & \textbf{0.1440 (0.2750) }\\
- & \checkmark  & - & - & 0.3865* (0.2677) & 0.1387* (0.1086) & 0.0985* (0.2717)& 0.7274* (0.1905) & 0.5549* (0.1876) & 0.1545* (0.2732)& 0.7108* (0.1754) & 0.5119* (0.1716) & 0.0842* (0.2745)& 0.7776 (0.1348) & 0.6001 (0.1769) & 0.1182* (0.2773)& \textbf{0.7848 (0.1489)} & \textbf{0.6120 (0.1747)} & \textbf{0.2158 (0.2718)} \\
- & - & \checkmark & - & 0.0000* (0.0000) & 0.0000* (0.0000) & 0.0824* (0.2749)& 0.5511* (0.2340) & 0.4363* (0.2221) & 0.6116* (0.3004)& 0.4599* (0.2394) & 0.3454* (0.2075) & 0.5931* (0.3255)& \textbf{0.6519 (0.2140)} & \textbf{0.5345* (0.2317)} &\textbf{ 0.7045 (0.2771)}& 0.6337 (0.2241) & 0.5096 (0.2363) & 0.6830 (0.2914) \\
- & - & - & \checkmark  & 0.0837* (0.1190) & 0.0749* (0.1039) & 0.1084* (0.2757)& 0.7854* (0.1417) & 0.6431* (0.1641) & 0.1482* (0.2809)& 0.7365* (0.1814) & 0.5465* (0.1543) & \textbf{0.2326 (0.2868)}& 0.8081* (0.1425) & 0.6933 (0.1688) & 0.1253* (0.2764)& \textbf{0.8265 (0.1236)} & \textbf{0.6981 (0.1590)} & 0.2285 (0.2836) \\ \hline

\checkmark  & \checkmark  & - & - & 0.0829* (0.1442) & 0.0167* (0.0324) & 0.0824* (0.2749)& 0.7476* (0.1850) & 0.5890* (0.1897) & 0.1476* (0.2819)& 0.7355* (0.1776) & 0.5771* (0.1918) & 0.1084* (0.2779)& \textbf{0.8133 (0.1262)} & \textbf{0.6642 (0.1623)} & 0.1204* (0.2911)& 0.8057 (0.1340) & 0.6543 (0.1682) & \textbf{0.1895 (0.2716)} \\
\checkmark  & - & \checkmark  & - & 0.0003* (0.0011) & 0.0002* (0.0008) & 0.0830* (0.2747)& 0.5983* (0.2102) & 0.4786* (0.2074) & 0.6653* (0.2896)& 0.2651* (0.2347) & 0.2091* (0.1976) & 0.4976* (0.3462)& \textbf{0.7052* (0.1960)} & \textbf{0.5807* (0.2212)} & \textbf{0.7395* (0.2566)}& 0.6591 (0.2251) & 0.5348 (0.2401) & 0.6962 (0.2889) \\
\checkmark  & - & - & \checkmark  & 0.5383* (0.2552) & 0.4540* (0.2097) & 0.1807 (0.2825)& 0.8352* (0.1245) & 0.6921* (0.1574) & 0.1497* (0.2838)& 0.8201* (0.1395) & 0.6423* (0.1531) & 0.1665* (0.2879)& 0.8541 (0.1212) & 0.7151 (0.1596) & 0.1069* (0.2809)& \textbf{0.8550 (0.1086)} & \textbf{0.7190 (0.1499)} & \textbf{0.2044 (0.2806)} \\
- & \checkmark  & \checkmark  & - & 0.3325* (0.2605) & 0.2110* (0.2008) & 0.4098* (0.3445)& 0.7659* (0.1621) & 0.6411* (0.1930) & 0.7369* (0.2377)& 0.8200 (0.1167) & 0.6983 (0.1619) & 0.7650 (0.2333)& \textbf{0.8285 (0.1085)} & \textbf{0.7107 (0.1662)} & 0.7748 (0.2235)& 0.8168 (0.1231) & 0.6963 (0.1825) & \textbf{0.7834 (0.2168)} \\
- & \checkmark  & - & \checkmark  & 0.8428* (0.1026) & 0.6758* (0.1491) & 0.2294* (0.2797)& 0.8487* (0.1051) & 0.6905* (0.1667) & 0.1551* (0.2801)& 0.8556* (0.1010) & 0.7023* (0.1537) & 0.2293 (0.2790)& 0.8470* (0.1105) & 0.6988* (0.1660) & 0.1293* (0.2768)&\textbf{ 0.8672 (0.0969)} & \textbf{0.7172 (0.1508)}& \textbf{0.2589 (0.2856)} \\
- & - & \checkmark  & \checkmark  & 0.2007* (0.1836) & 0.1912* (0.1748) & 0.1566* (0.2941)& 0.8488* (0.1070) & 0.7283* (0.1675) & 0.7530* (0.2290)& 0.8293* (0.1235) & 0.6947* (0.1622) & 0.7464* (0.2271)& \textbf{0.8765 (0.0886)} & \textbf{0.7689 (0.1488)} & 0.7711 (0.2273)& 0.8715 (0.0981) & 0.7668 (0.1603) & 0\textbf{.7742 (0.2264)} \\ \hline

\checkmark  & \checkmark  & \checkmark  & - & 0.2679* (0.2380) & 0.1733* (0.1925) & 0.4505* (0.3586)& 0.7754* (0.1649) & 0.6588* (0.1963) & 0.7437* (0.2491)& 0.7653* (0.1865) & 0.6513* (0.2107) & 0.7533 (0.2575)& \textbf{0.8338* (0.1072)} & \textbf{0.7237* (0.1674)} & \textbf{0.7890 (0.2136)}& 0.8228 (0.1226) & 0.7105 (0.1805) & 0.7842 (0.2166) \\
\checkmark  & \checkmark  & - & \checkmark  & 0.8503* (0.1044) & 0.6962* (0.1454) & 0.2376 (0.2788)& 0.8614* (0.1031) & 0.7189* (0.1518) & 0.1529* (0.2862)& 0.8632* (0.0974) & 0.7011* (0.1475) & 0.2128 (0.2767)& 0.8587* (0.1080) & 0.7184 (0.1548) & 0.1168* (0.2854)& \textbf{0.8714 (0.0952)} & \textbf{0.7289 (0.1449)} & \textbf{0.2207 (0.2803)} \\
\checkmark  & - & \checkmark  & \checkmark  & 0.4983* (0.2883) & 0.4325* (0.2733) & 0.5341* (0.3453)& 0.8610* (0.1114) & 0.7584* (0.1602) & 0.7751 (0.2276)& 0.8566* (0.1104) & 0.7502* (0.1597) & 0.7805 (0.2312)& \textbf{0.8793* (0.0974)} & \textbf{0.7797* (0.1501}) & \textbf{0.7846* (0.2264)}& 0.8697 (0.1045) & 0.7700 (0.1567) & 0.7785 (0.2304) \\
- & \checkmark  & \checkmark  & \checkmark  & 0.8620* (0.0953) & 0.7500* (0.1592) & 0.7607* (0.2281)& 0.8766* (0.0941) & 0.7626* (0.1589) & 0.7839* (0.2092)& 0.8769* (0.0965) & 0.7686* (0.1576) & 0.7785* (0.2117)& 0.8812 (0.0924) & 0.7735 (0.1526) & 0.7792* (0.2189)& \textbf{0.8832 (0.0935)} & \textbf{0.7794 (0.1558)} & \textbf{0.7918 (0.2101)} \\ \hline\hline
\multicolumn{4}{c|}{Average} & 0.6196* (0.3202) & 0.5130* (0.3041) & 0.4957* (0.3598)& 0.8436* (0.1279) & 0.7247* (0.1728) & 0.6139* (0.3619)& 0.8405* (0.1356) & 0.7178* (0.1766) & 0.6313* (0.3446)&\textbf{ 0.8632 (0.1033)} & \textbf{0.7488 (0.1588)} & 0.6174* (0.3743)& \textbf{0.8618 (0.1071)} & \textbf{0.7472 (0.1625)} & \textbf{0.6438 (0.3397)} \\

% \checkmark & \checkmark & \checkmark & \checkmark & 0.8819 / 0.7818 / 0.7908 & -\quad /\quad -\quad /\quad - & -\quad /\quad -\quad /\quad - \\

\hline
\end{tabular}
}
\end{center}

\end{table*}
%%%%%%%%%%%%%%%%%%%%%%%%%%%%

\textbf{IXI Dataset} consists of healthy brain MRI scans from three different institutions, each with three MR sequences: T1, T2, and PD-weighted (PD). The dataset includes 576 subjects in total; we divided them into training, validation, and test sets of 270, 30, and 276 subjects, respectively. In the preprocessing step, we first apply an affine transformation to the T1 and PD images to align them with the T2 images, by utilizing the NMI similarity metric through the use of the greedy registration tool~\citep{yushkevich2016ic}. The original intensities of the images are linearly adjusted to a range of $[-1, 1]$ with 99.5 percentile as the upper bound. During training, 2D axial slices are used, with 80 center axial slices specifically selected. The size of the final images is set to $256 \times 256$. The number of slices used for the training, validation, and test sets is 21,600, 2,400, and 22,080, respectively.

\subsection{Competing Methods}
We compare our unified framework with several state-of-the-art MR image synthesis methods, including TransMI~\citep{cho2024disentangled}, MMGAN~\citep{sharma2019missing}, ResViT~\citep{dalmaz2022resvit}, and Hyper-GAE~\citep{yang2023learning}. TransMI is designed for one-to-one translation with a unified approach to use a single generator for all of the modality pairs. However, it can only synthesize $N$ cases out of $2^N-2$ missing cases, so we extend this work $\textrm{TransMI}_{uni}$ to many-to-one using the multichannel early fusion encoder, while maintaining the unified approach. MMGAN, a foundational FCN-based framework; ResViT, a well-regarded transformer-based model; and Hyper-GAE, a recent, fully 3D graph-based approach. For Hyper-GAE, we train with 3D images and evaluate with 2D slices. The results of all methods were produced using public code repositories~\footnote{https://github.com/trane293/mm-gan}~\footnote{https://github.com/icon-lab/ResViT}~\footnote{https://github.com/HeranYang/hyper-GAE}. We evaluate all missing scenarios with 14 cases and 28 synthesis results for the BraTS dataset and 6 cases and 9 synthesis results for the IXI dataset, using two standard metrics, peak signal-to-noise ratio (PSNR) and structural similarity measure (SSIM). For the downstream clinical task of data imputation, we utilized the Dice similarity coefficient to measure the accuracy of tumor segmentation. The statistical significance of our results was confirmed using the Wilcoxon signed-rank test, with a threshold of p $<$ 0.05.

\subsection{Multisequence MRI synthesis} 
\label{synthresult}
\subsubsection{Synthesis on Pathological Data (BraTS)}
The BraTS dataset is particularly challenging due to the high variability in tumor shape, size, and location, making it a robust testbed for a model's ability to handle complex, clinically relevant scenarios. On this dataset, our proposed HF-GAN demonstrates a clear and statistically significant advantage.
As detailed in~\autoref{table:quan_brats}, HF-GAN achieves the highest overall average PSNR of 30.27 and SSIM of 0.9393. This represents a notable improvement over the next-best performing methods, the 3D Hyper-GAE (PSNR 29.54, SSIM 0.9361) and our extended baseline TransMI (PSNR 29.45, SSIM 0.9305). This superior performance is a direct result of our hybrid-fusion architecture, which is constructed to extract and integrate the crucial complementary information that defines tumor boundaries and internal structures. This information is often challenging for single-pathway encoders to synthesize accurately.
The qualitative results in~\autoref{fig:qual_brats} provide compelling visual evidence of this advantage. As shown in the first row, ResViT faces challenges with the latent space. It struggles to synthesize tumor regions when the information provided is scarce, while it achieves great synthesis results when sufficient information is available. Hyper-GAE also yields a fuzzy and distorted image of the hyper-intense tumor. Conversely, HF-GAN offers a significantly clearer depiction of the pathology that aligns more closely with the actual ground truth. This demonstrates a remarkable capacity to deduce intricate pathological features, even with sparse source data.

\begin{figure*}[!t]
\begin{center}
\includegraphics[width=0.95\textwidth]{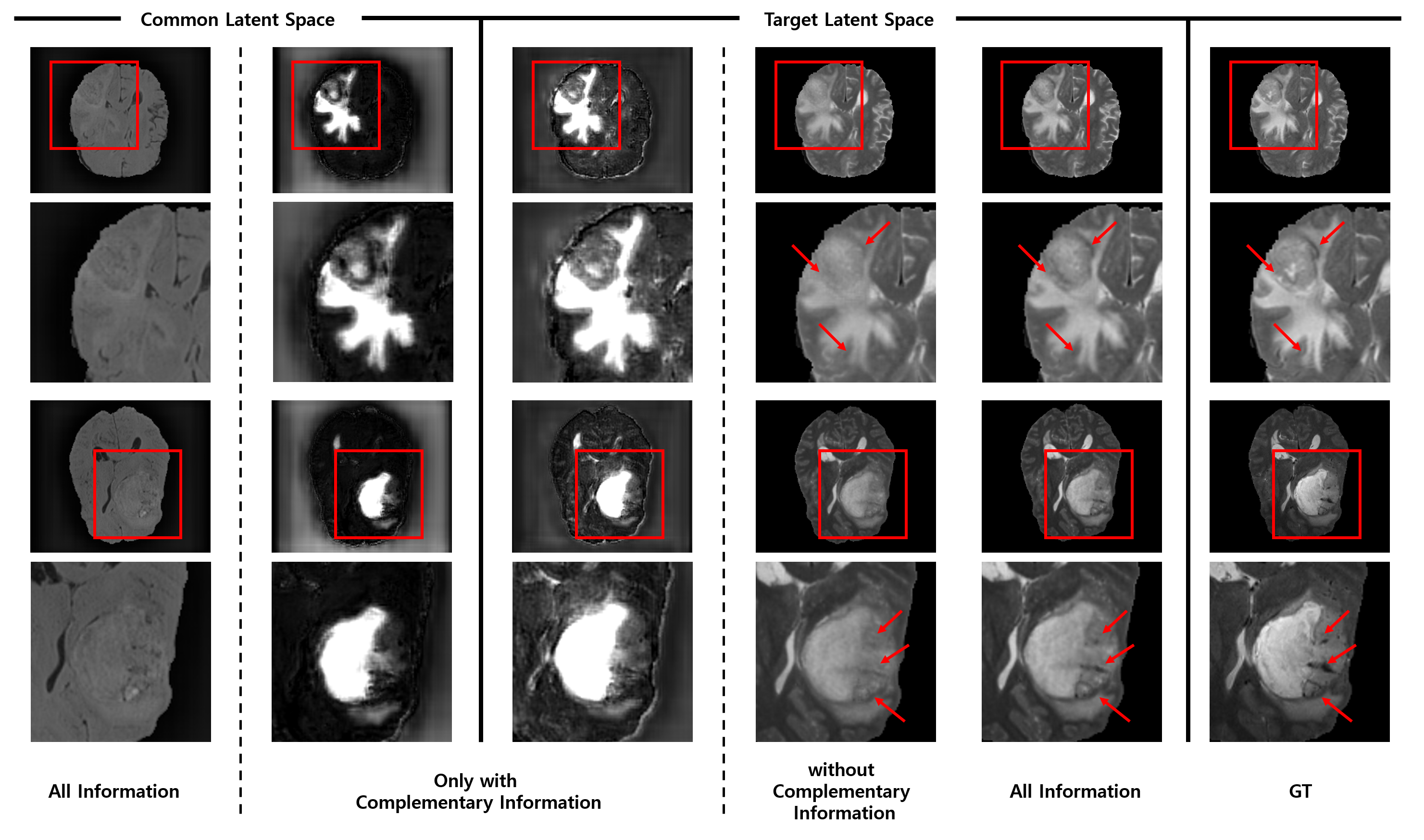}
\end{center} 
\caption{Image synthesis results of different feature representations when the same MR images are used. The feature representations in the common latent space are used to synthesize the images in the first and second columns, while the feature representations in the target (T2) latent space are used to synthesize the images in the third to fifth columns. Each column is synthesized from three distinct feature representations: 1) using all extracted features (all information), 2) using only features from the early fusion encoder (only with complementary information), and 3) using only features from the modality-specific encoders (without complementary information).} 
\label{fig:analysis}
\end{figure*} 

\subsubsection{Synthesis on Healthy Data (IXI)}
On the IXI dataset of healthy brains, where the synthesis task depends more on fine anatomical detail than on overt pathology, the performance gap between methods naturally narrows. Even in this context, HF-GAN remains the top-performing model, as shown in~\autoref{table:quan_ixi}. Our method achieves the highest average PSNR (28.89) and SSIM (0.9110), again showing a statistically significant lead over all competitors. While the numerical gains are more modest here, the consistent advantage demonstrates the fundamental strength of our design. It suggests that the principled disentanglement and fusion of features benefit not only the synthesis of pathology but also the precise rendering of complex anatomical structures, such as the boundaries between gray and white matter. You can see outstanding synthesized results in~\autoref{fig:qual_ixi}.

\subsection{Data Imputation for Segmentation}
The goal of a medical image synthesis model is its utility in clinical workflows. We tested this by using the generated images to fill missing sequences in a dataset before feeding it to a brain tumor segmentation network~\citep{isensee2021nnu}. Results are summarized in~\autoref{table:seg}. 

The most critical result is the comparison against the "Without Synthesis" baseline, where attempting segmentation with incomplete data leads to a dramatic failure in the model's predictive ability. For instance, if only the T2 sequence is present, the Whole Tumor Dice score is limited to just 0.3865. However, by using HF-GAN to impute the missing data, this score rises significantly to 0.7848, representing an improvement of 40 absolute percentage points.
Furthermore, when comparing imputation methods, our model demonstrates highly competitive results against Hyper-GAE, especially considering the architectural differences. Our proposed method operates on a 2D slice-by-slice basis, whereas Hyper-GAE leverages a fully 3D volumetric approach that has an inherent advantage in utilizing inter-slice spatial context. Despite this, our results for Whole Tumor (WT) (0.8618 compared to 0.8632) and Tumor Core (TC) (0.7472 compared to 0.7488) are comparable to those of Hyper-GAE, showing only marginal differences. Crucially, our method's strength is evident in segmenting the most challenging sub-region, the Enhanced Tumor (ET), where we achieve a Dice score of 0.6438, significantly outperforming Hyper-GAE's score of 0.6174. This indicates that our method successfully restores important intra-slice features, serving as a strong and efficient option, particularly excelling in scenarios where detailed precision is crucial.

\begin{figure}[!t]
\begin{center}
\includegraphics[width=0.8\columnwidth]{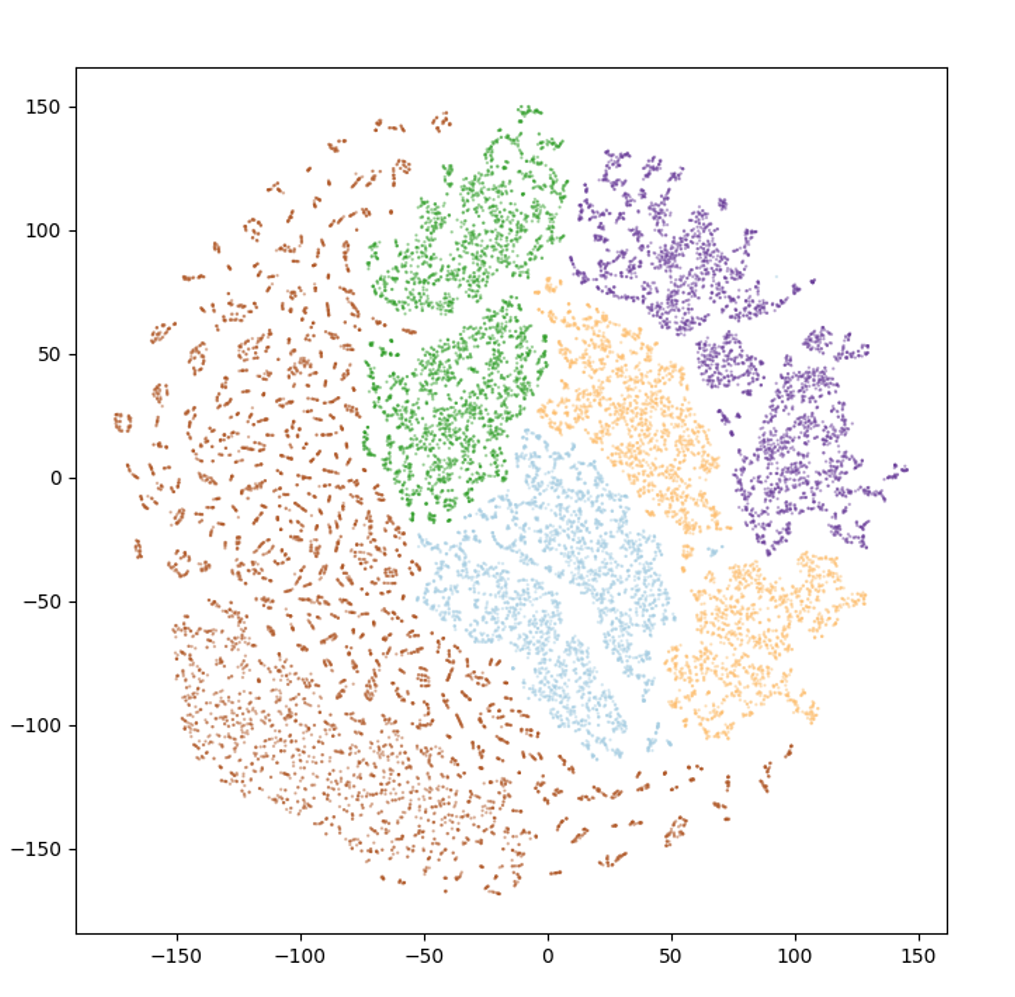}
\end{center} 
\caption{T-SNE visualization of common latent space and target latent space. A total of 21,000 feature representations are depicted, covering all 14 missing scenarios. The brown color illustrates the common feature representations, while the other colors indicate the feature representations within the target latent spaces transformed by the modality infuser. T1 is shown in blue, T2 in green, T1c in yellow, and FL in purple.} 
\label{fig:tsne}
\end{figure} 

%%%%%%%%%%%%%%%%%%%%%%%%%%%%%
\begin{table}[!b]
\caption{\label{table:ablation1} Evaluation results of our method and its variants, with the hybrid-fusion encoder and channel attention-based feature fusion ablated. The results are averaged over all missing scenarios.}
\begin{center}
\resizebox{1.0\columnwidth}{!}{
\begin{tabular}{>{\raggedright}p{40mm}|>{\centering}p{15mm}>{\centering\arraybackslash}p{15mm}}
\hline 
\multirow{2}{*}{Methods} & \multicolumn{2}{c}{Results}  \\ \cline{2-3}
 & PSNR & SSIM \\
\hline

Ours (w/o $Enc^{M}$) & 29.19 & 0.9333 \\
Ours (w/o $Enc^{C}$) & 29.41 & 0.9324 \\
Ours (w/o $CAFF$) & 29.40 & 0.9331 \\
Ours &  \textbf{29.67} & \textbf{0.9354} \\

\hline
\end{tabular}
}
\end{center}
\end{table}

\begin{figure}[!b]
\begin{center}
\includegraphics[width=1.0\columnwidth]{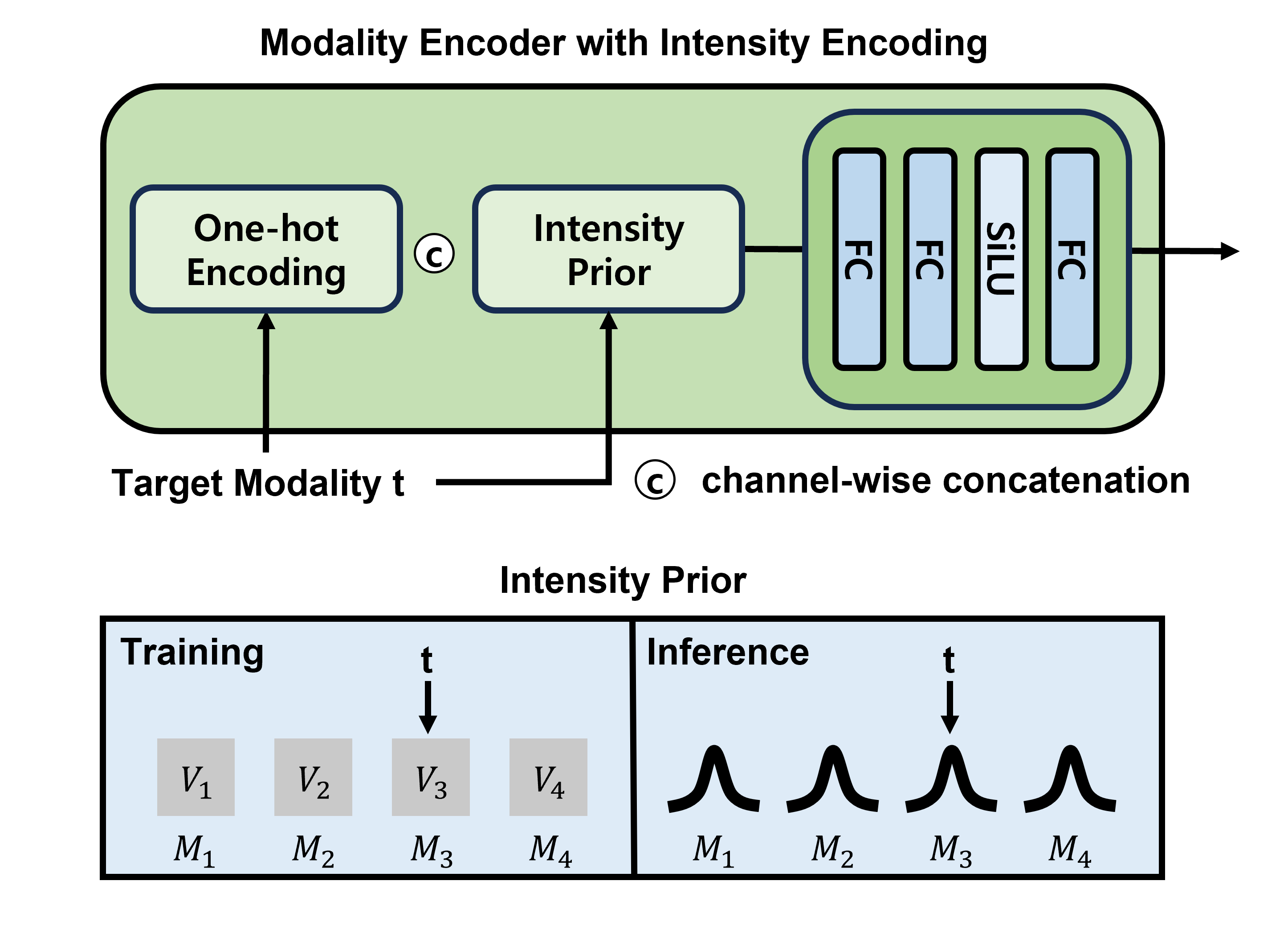}
\end{center} 
\caption{The structure of the intensity encoding module $IE$. The intensity prior is pre-computed based on the training data.} 
\label{fig:ie}
\end{figure} 

\subsection{Analysis of Complementary Information and Latent Space}
\label{sec:analysis}
Our framework is designed around a core principle: the separation of universal anatomical information from modality-specific appearance. This is achieved through a disentangled representation managed by a hybrid-fusion encoder and a partitioned latent space. The encoder’s primary function is to extract two distinct types of information from the available input sequences: modality-specific features, which capture the unique contrast of each MR sequence, and complementary features, which represent synergistic details that emerge only when multiple modalities are available. These features are then organized into a common latent space, intended to store the modality-agnostic anatomical map, and a target latent space, which holds the contrast for the desired output image.

The key to our model's effectiveness lies in how it constructs the common latent space. The encoder first aggregates the input modalities to derive the rich, complementary information, which is particularly effective at identifying complex tissue characteristics such as tumor boundaries or edema. This complementary information is not the anatomical map itself; rather, it is used as crucial evidence to refine it. For example, by learning from T2 and FLAIR scans that a certain region is edema, the model can delineate its structural boundaries with high precision. This refined boundary—a purely anatomical feature—is then encoded into the common latent space. This ensures the model’s foundational anatomical map is not a simple average, but an accurate representation informed by cross-modal insights.

Visual analysis in~\autoref{fig:analysis} confirms that this process works as intended.
When an image is synthesized using only the common latent space, the output preserves clear anatomical structures but lacks any specific MR contrast. This validates its role as the universal anatomical map.
Conversely, visualizing the contribution of the early fusion features specifically highlights pathological regions, confirming that this is where the most critical complementary information is found.
When complementary features are withheld from the synthesis process, the resulting images exhibit a noticeable loss of quality, particularly in the clarity and definition of tumor boundaries. This outcome demonstrates that the complementary features are essential for constructing an accurate structural representation in the common latent space. 

Furthermore, the t-SNE plot in~\autoref{fig:tsne} shows a distinct separation between the feature clusters, providing quantitative evidence that the model successfully disentangles different types of latent space.

% \begin{table}[!t]
% \caption{\label{table:ablation1} Evaluation results of our method and its variants, with the hybrid-fusion encoder and channel attention-based feature fusion ablated. The results are averaged over all missing scenarios.}
% \begin{center}
% \resizebox{1.0\columnwidth}{!}{
% \begin{tabular}{>{\raggedright}p{40mm}|>{\centering}p{15mm}>{\centering}p{15mm}>{\centering}p{15mm}>{\centering\arraybackslash}p{15mm}}
% \hline 
% \multirow{2}{*}{Methods} & \multicolumn{4}{c}{Results}  \\ \cline{2-5}
%  & MAE & PSNR & SSIM & MS-SSIM \\
% \hline

% Ours (w/o $Enc^{MF}$) & 0.0151 & 29.19 & 0.9333 & 0.9446 \\
% Ours (w/o $Enc^{EF}$) & 0.0144 & 29.41 & 0.9324 & 0.9459 \\
% Ours (w/o $CAFF$) & 0.0145 & 29.40 & 0.9331 & 0.9456 \\
% Ours & \textbf{0.0139} & \textbf{29.67} & \textbf{0.9354} & \textbf{0.9486} \\

% \hline
% \end{tabular}
% }
% \end{center}
% \end{table}

\begin{figure*}[!t]
\begin{center}
\includegraphics[width=0.95\textwidth]{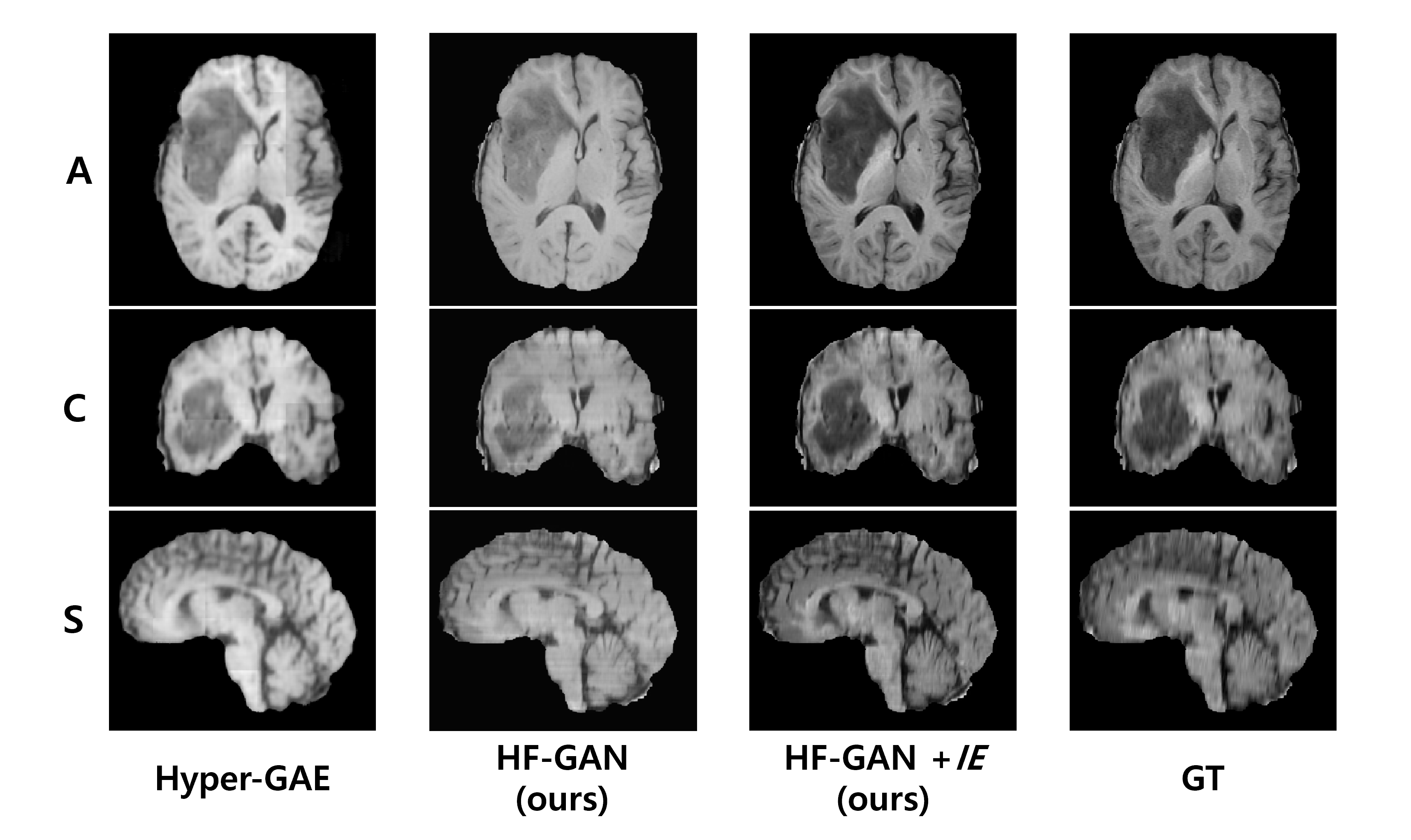}
\end{center} 
\caption{An example of 3D synthesized results for the missing MR sequence on the BraTS dataset. A, C, and S represent axial, coronal, and sagittal views, respectively.} 
\label{fig:3d_qual}
\end{figure*} 

\subsection{Ablation Study}

To verify the effectiveness of our proposed framework, we perform an ablation study. \autoref{table:ablation1} presents the evaluation results with different compositions. Removing either the modality-specific encoders ($Enc^M$) or the complementary encoder ($Enc^C$) results in a significant drop in performance, with PSNR falling to 29.19 and 29.41, respectively. This confirms that both feature types are essential and that extracting them through specialized pathways is beneficial. The necessity of a dynamic fusion strategy is confirmed by ablating the CAFF module. Replacing it with a simpler fusion method caused the PSNR to drop to 29.40, validating the contribution of our attention-based integration for effectively combining feature streams.

%%%%%%%%%%%%%%%%%%%%%%%%%%%%%
% \begin{table}[t]
% \caption{\label{table:ablation2} Evaluation results of the modality encoding methods of the modality infuser. The results are averaged over all missing scenarios.}
% \begin{center}
% \resizebox{1.0\columnwidth}{!}{
% \begin{tabular}{>{\raggedright}p{40mm}|>{\centering}p{15mm}>{\centering\arraybackslash}p{15mm}}
% \hline 
% \multirow{2}{*}{Modality Encoding} & \multicolumn{2}{c}{Results}  \\ \cline{2-3}
% & PSNR & SSIM \\
% \hline

% learnable parameter & 26.79 & 0.8986 \\
% sinusoidal function & 28.81 & 0.9258 \\
% one-hot vector + MLP & \textbf{29.63} & \textbf{0.9353} \\

% \hline
% \end{tabular}
% }
% \end{center}
% \end{table}

% \begin{table}[b]
% \caption{\label{table:ablation2} Evaluation results of the modality encoding methods of the modality infuser. The results are averaged over all missing scenarios.}
% \begin{center}
% \resizebox{1.0\columnwidth}{!}{
% \begin{tabular}{>{\raggedright}p{40mm}|>{\centering}p{15mm}>{\centering}p{15mm}>{\centering}p{15mm}>{\centering\arraybackslash}p{15mm}}
% \hline 
% \multirow{2}{*}{Modality Encoding} & \multicolumn{4}{c}{Results}  \\ \cline{2-5}
%  & MAE & PSNR & SSIM & MS-SSIM \\
% \hline

% learnable parameter & 0.0194 & 26.79 & 0.8986 & 0.8861 \\
% sinusoidal function & 0.0153 & 28.81 & 0.9258 & 0.9381  \\
% one-hot vector + MLP & 0.0140 & 29.63 & 0.9353 & 0.9482 \\
% sinusoidal function + MLP & \textbf{0.0139} & \textbf{29.67} & \textbf{0.9354} & \textbf{0.9486} \\

% \hline
% \end{tabular}
% }
% \end{center}
% \end{table}

\section{Efficient High-Quality 3D Volume Synthesis}

%%%%%%%%%%%%%%%%%%%%%%%%%%%%%
\begin{table}[!b]
\caption{\label{table:3d} Quantitative evaluation results for 3D multisequence MRI synthesis on the BraTS dataset. $mean$ represents the statistical mean of intensity prior of the training dataset, and $GT$ represents the ground truth intensity prior. The results are averaged over all missing scenarios.}
\begin{center}
\resizebox{1.0\columnwidth}{!}{
\begin{tabular}{>{\raggedright}p{40mm}|>{\centering}p{15mm}>{\centering\arraybackslash}p{15mm}}
\hline 
\multirow{2}{*}{Methods} & \multicolumn{2}{c}{Results}  \\ \cline{2-3}
& PSNR & SSIM \\
\hline

Hyper-GAE (3D) & 29.54 & 0.9361 \\
HF-GAN (ours) & 30.27 & 0.9393 \\
HF-GAN + $IE$ ($mean$) & \textbf{30.62} & \textbf{0.9434} \\ \hline
HF-GAN + $IE$ ($GT$) & \textbf{32.45} & \textbf{0.9502} \\
\hline
\end{tabular}
}
\end{center}
\end{table}

A fundamental dilemma in volumetric medical image synthesis is the trade-off between 2D and 3D architectures. Fully 3D methods, such as Hyper-GAE, process entire volumes at once, inherently ensuring perfect anatomical consistency across all slices. However, this comes at a high computational cost, requiring substantial GPU memory that is often impractical for many research and clinical environments. Conversely, 2D slice-based methods are computationally efficient and can be trained on standard hardware. Their critical drawback, however, is a lack of consistency between adjacent slices, as each slice is processed independently. This can lead to unrealistic "stripe" artifacts when the 2D slices are stacked into a 3D volume and viewed from a different axis.

Our framework is designed to provide a practical, high-performance solution to this problem. We demonstrate that a well-designed 2D architecture can not only mitigate consistency issues but also achieve a synthesis quality that rivals or even exceeds that of more demanding 3D models. To do this, we enhance our adaptable framework with an Intensity Encoding (IE) module. This lightweight module, integrated into the $MI$ as shown in~\autoref{fig:ie}, conditions the synthesis of each slice on a consistent intensity prior learned from the target volume. This simple but effective mechanism explicitly enforces the slice-to-slice intensity consistency that standard 2D methods lack, without incurring the heavy cost of full 3D convolutions. We used the median value of the soft tissue region as an intensity prior in this work.

The results of this approach are compelling. As demonstrated  in~\autoref{table:3d} and \autoref{fig:3d_qual}, our enhanced 2D framework with $IE$ achieves the highest quantitative scores, while also exhibiting a high level of precision in the depiction of fine details, thus ensuring the absence of any stripe artifacts. It is noteworthy that the baseline 2D HF-GAN demonstrates superior performance in comparison to the fully 3D Hyper-GAE framework, as evidenced by higher PSNR (30.62 compared to 29.54) and SSIM (0.9434 compared to 0.9361). This finding indicates that a sophisticated 2D architecture may prove more effective than a complex 3D model, particularly in scenarios where hardware constraints limit the training of a more extensive network. 
The proposed methodology maintains the efficiency of a two-dimensional workflow while introducing three-dimensional-aware consistency through the $IE$ module. This approach offers a "best of both worlds" solution. The first-place result~\citep{cho2024twostageapproachbrainmr} in ASNR-MICCAI BraTS MRI Synthesis Challenge 2024 (BraSyn)~\citep{bonato2025advancing} further validates the assertion that this efficient and adaptable approach represents the state-of-the-art in MRI synthesis.

\section{Discussion and Conclusion}
In this work, we propose a unified synthesis framework for multisequence MRI. We design a hybrid-fusion encoder that facilitates the extraction of complementary information, coupled with a feature fusion module that uses channel attention to construct a robust common latent space and a modality infuser that enables an efficient synthesis of the target sequence. The experimental results demonstrate that our proposed method can achieve state-of-the-art performance in multisequence MRI synthesis and show potential for application in data imputation. A further investigation is conducted into the impact of the components that have been designed, and an ablation study is performed to confirm their effectiveness.

While our method shows superior results both quantitatively and qualitatively, we note that there are still certain limitations. The use of multiple encoders to construct a common latent space can result in high memory consumption, especially when dealing with a large number of modalities. As in our previous work~\citep{cho2024disentangled}, the use of a weight-shared encoder could be a potential solution, although it involves a trade-off between performance and memory efficiency. Our framework is specifically designed for spatially registered datasets, but it requires additional time for atlas-based image registration. Nevertheless, our framework has the potential to be applied to unpaired datasets, since our training loss includes both the reconstruction loss for spatially aligned images and the cycle-consistency loss applicable to unpaired datasets. Therefore, extending our approach to unpaired datasets, including translation between MRI and CT, is part of our future work. 

Furthermore, our analysis confirms the significant benefit of the proposed disentanglement approach, which effectively isolates and leverages different feature types. We believe this architectural principle is not limited to the GAN-based framework presented in this work; it could be integrated into other advanced generative backbones, such as diffusion models or state-space models like Mamba, to further advance the field of medical image synthesis.

\section*{Acknowledgments}
This work was supported by Institute for Information \& communications Technology Promotion(IITP) grant funded by the Korea government (MSIT) (No.00223446, Development of object-oriented synthetic data generation and evaluation methods).

%%Harvard
\bibliographystyle{model2-names.bst}\biboptions{authoryear}
\bibliography{main}

\end{document}